\documentclass{tlp}

\usepackage{aopmath}
\usepackage[latin1]{inputenc}
\usepackage{psfig}
\usepackage{latexsym}
\usepackage{oz2tex}

\newcommand{\nt}[1]{$\langle\textsf{#1}\rangle$}

\begin{document}


\title[Logic programming in Oz]{Logic programming in the context of\\
multiparadigm programming:\\
the Oz experience \footnote{
This article is a much-extended version of the tutorial talk
``Logic Programming in Oz with Mozart''
given at the International Conference on Logic Programming,
Las Cruces, New Mexico, Nov.~1999.
Some knowledge of traditional logic programming
(with Prolog or concurrent logic languages) is assumed.}
}

\author[P. Van Roy et al.]
{PETER VAN ROY\\
Universit\'{e} catholique de Louvain,
B-1348 Louvain-la-Neuve, Belgium
\and
 PER BRAND\\
Swedish Institute of Computer Science, S-164 28 Kista, Sweden
\and
 DENYS DUCHIER\\
Universit\"at des Saarlandes, D-66123 Saarbr\"ucken, Germany
\and
 SEIF HARIDI\\
Royal Institute of Technology (KTH), S-164 28 Kista, Sweden
\and
 MARTIN HENZ\\
National University of Singapore, Singapore 117543
\and
 CHRISTIAN SCHULTE\\
Royal Institute of Technology (KTH), S-164 28 Kista, Sweden}


\maketitle
\label{firstpage}
 
\begin{abstract}
Oz is a multiparadigm language that supports logic programming as one
of its major paradigms.  A multiparadigm language is designed to
support different programming paradigms (logic, functional,
constraint, object-oriented, sequential, concurrent, etc.) with equal
ease.  This article has two goals: to give a tutorial of logic
programming in Oz and to show how logic programming fits naturally
into the wider context of multiparadigm programming.  
Our experience shows that there are two classes of problems, which we
call {\em algorithmic} and {\em search} problems, for which logic
programming can help formulate practical solutions.
{\em Algorithmic} problems have known efficient algorithms.
{\em Search} problems do not have
known efficient algorithms but can be solved with search.
The Oz support for logic programming targets these two
problem classes specifically, using the concepts needed for each.
This is in contrast to the Prolog approach, which targets
both classes with one set of concepts, which results in
less than optimal support for each class.
We give examples
that can be run interactively on the Mozart system,
which implements Oz.  
To explain the essential difference between algorithmic and
search programs, we define the Oz execution model.
This model subsumes both concurrent
logic programming (committed-choice-style) and search-based logic
programming (Prolog-style).  Furthermore, as consequences of
its multiparadigm nature, the model supports new abilities
such as first-class top levels, deep guards, active objects, and
sophisticated control of the search process.  Instead of Horn clause
syntax, Oz has a simple, fully compositional, higher-order syntax that
accommodates the abilities of the language.  
We give a brief
history of Oz that traces the development of its main ideas
and we summarize the lessons learned from this work.
Finally, we give many entry points into the Oz literature.
\end{abstract}

\section{Introduction}

In our experience,
logic programming can help give
practical solutions to many different problems.
We have found that all these problems
can be divided into two classes,
each of which uses a totally different approach:
\begin{itemize}
\item {\bf Algorithmic problems}.
These are problems for which efficient algorithms are known.
This includes parsing, rule-based expert systems,
and transformations of complex symbolic data structures.
For such problems,
a logical specification of the algorithm
is sometimes simpler than an imperative specification.
In that case, {\em deterministic logic programming}
or {\em concurrent logic programming} may be natural ways
to express it.
Logical specifications are not always simpler.
Sometimes an imperative specification is better,
e.g., for problems in which
state updating is frequent.
Many graph algorithms are of the latter type.

\item {\bf Search problems}.
These are problems for which efficient algorithms are not known.
This may be either because such algorithms are
not possible in principle
or because such algorithms have not been made explicit.
We cite, e.g., NP-complete problems~\cite{Garey.79}
or problems with complex specifications
whose algorithms are difficult for this reason.
Some examples of search problems are
optimization problems (planning, scheduling, configuration),
natural language parsing, and theorem proving.
These kinds of problems can be solved by doing search, i.e.,
with {\em nondeterministic logic programming}.
But search is a dangerous tool.
If used naively,
it does not scale up to real applications.
This is because the size of the search space grows
exponentially with the problem size.
For a real application, all possible effort must be
made to reduce the need for search: use strong (global) constraints,
concurrency for cooperative constraints,
heuristics for the search tree, etc.~\cite{fdtutorial}.
For problems with complex specifications,
using sufficiently strong constraints
sometimes results in a polynomial-time algorithm~\cite{CP-NL}.
\end{itemize}
For this paper, we consider
{\em logic programming} as programming with
executable specifications written in a simple logic
such as first-order predicate calculus.
The Oz support for logic programming is targeted
specifically towards the two
classes of algorithmic problems and search problems.
The first part of this article
(Sections~\ref{detlp}--\ref{moresearch})
shows how to write logic programs
in Oz for these problems.
Section~\ref{detlp} introduces deterministic logic programming,
which targets algorithmic problems.
It is the most simple and direct way of doing logic programming in Oz.
Section~\ref{nondetlp} shows how to do
nondeterministic logic programming in the Prolog style.
This targets neither algorithmic nor search problems,
and is therefore only of pedagogical interest.
Section~\ref{conclp} shows how to do concurrent logic
programming in the classical tradition.
This targets more algorithmic problems.
Section~\ref{state} extends Section~\ref{conclp}
with state.
In our experience, state
is essential for practical concurrent logic programming.
Section~\ref{moresearch}
expands on Section~\ref{nondetlp} to show how search
can be made practical.

The second part of this article
(Sections~\ref{executionmodel}--\ref{lessons})
focuses on the essential difference between the
techniques used to solve
algorithmic and search problems.
This leads to the wider context
of {\em multiparadigm} programming.
Section~\ref{executionmodel} introduces the Oz execution model,
which has a strict functional core and
extensions for concurrency, lazy evaluation,
exception handling, security, state, and search.
The section explains how these extensions can be
used in different combinations
to provide different programming paradigms.
In particular, Section~\ref{compspaces} explains the abstraction
of {\em computation spaces},
which is the main tool for doing search in Oz.
Spaces make possible a deep synthesis of concurrent and
constraint logic programming.
Section~\ref{related} gives an overview of other
research in multiparadigm programming
and a short history of Oz.
Finally,
Section~\ref{lessons} summarizes the lessons we have learned in
the Oz project on how to do practical logic programming
and multiparadigm programming.

This article gives an informal (yet precise)
introduction targeted towards Prolog programmers.
A more complete presentation
of logic programming in Oz and
its relationship to other programming concepts
is given in the textbook~\cite{book}.

\section{Deterministic logic programming}
\label{detlp}

We call {\em deterministic} logic programming the case
when the algorithm's control
flow is completely known and specified by the programmer.
No search is needed.
This is perfectly adapted to sequential algorithmic problems.
For example, a deterministic naive reverse can be
written as follows in Oz:
\begin{oz2texdisplay}\OzSpace{3}\OzKeyword{declare}\OzEol
\OzSpace{3}\OzKeyword{proc}\OzSpace{1}\OzChar\{Append\OzSpace{1}Xs\OzSpace{1}Ys\OzSpace{1}Zs\OzChar\}\OzEol
\OzSpace{6}\OzKeyword{case}\OzSpace{1}Xs\OzEol
\OzSpace{6}\OzKeyword{of}\OzSpace{1}nil\OzSpace{1}\OzKeyword{then}\OzSpace{1}Zs=Ys\OzEol
\OzSpace{6}[]\OzSpace{1}X|Xr\OzSpace{1}\OzKeyword{then}\OzSpace{1}Zr\OzSpace{1}\OzKeyword{in}\OzEol
\OzSpace{9}Zs=X|Zr\OzSpace{1}\OzChar\{Append\OzSpace{1}Xr\OzSpace{1}Ys\OzSpace{1}Zr\OzChar\}\OzEol
\OzSpace{6}\OzKeyword{end}\OzEol
\OzSpace{3}\OzKeyword{end}\OzEol
\OzEol
\OzSpace{3}\OzKeyword{proc}\OzSpace{1}\OzChar\{NRev\OzSpace{1}Xs\OzSpace{1}Ys\OzChar\}\OzEol
\OzSpace{6}\OzKeyword{case}\OzSpace{1}Xs\OzEol
\OzSpace{6}\OzKeyword{of}\OzSpace{1}nil\OzSpace{1}\OzKeyword{then}\OzSpace{1}Ys=nil\OzEol
\OzSpace{6}[]\OzSpace{1}X|Xr\OzSpace{1}\OzKeyword{then}\OzSpace{1}R\OzSpace{1}\OzKeyword{in}\OzEol
\OzSpace{9}\OzChar\{NRev\OzSpace{1}Xr\OzSpace{1}R\OzChar\}\OzEol
\OzSpace{9}\OzChar\{Append\OzSpace{1}R\OzSpace{1}[X]\OzSpace{1}Ys\OzChar\}\OzEol
\OzSpace{6}\OzKeyword{end}\OzEol
\OzSpace{3}\OzKeyword{end}\end{oz2texdisplay}
This syntax should be vaguely familiar
to people with some knowledge of Prolog and functional
programming~\cite{artofprolog,maierwarren,haskellbook,caml}.
We explain it briefly, pointing out where it differs from Prolog.
All capitalized identifiers refer to logic variables
in a constraint store.\footnote{Including
\OzInline{Append} and \OzInline{NRev}, which are bound to procedure values
(lexically-scoped closures).}
\OzInline{Append} and \OzInline{NRev} are procedures whose arguments are passed
by unification, as in Prolog.
The \OzInline{\OzKeyword{declare}} declares new global identifiers,
\OzInline{Append} and \OzInline{NRev},
which are bound to the newly-created procedure values.
This means that the order of the declarations does not matter.
All local variables must be declared within a scope, e.g.,
``\OzInline{Zr\OzSpace{1}\OzKeyword{in}}'' and ``\OzInline{R\OzSpace{1}\OzKeyword{in}}'' declare \OzInline{Zr} and \OzInline{R}
with scopes to the next enclosing \OzInline{\OzKeyword{end}} keyword.
A list is either the atom \OzInline{nil} or a pair
of an element \OzInline{X} and a list \OzInline{Xr}, written \OzInline{X|Xr}.
The \OzInline{[]} is not an empty list,
but separates clauses in a \OzInline{\OzKeyword{case}} statement
(similar to a guarded command, except that \OzInline{\OzKeyword{case}} is sequential).

We explain briefly the semantics of the naive reverse
to highlight the relationship to logic programming.
The constraint store consists of equality constraints over
rational trees, similar to what is provided by many modern Prolog systems.
Statements are executed sequentially.
There are two basic operations on the store,
ask and tell~\cite{SaraswatBook}:
\begin{itemize}
\item The tell operation (e.g., \OzInline{Ys=nil})
adds a constraint; it performs unification.
The tell is an {\em incremental tell}; if the constraint
is inconsistent with the store then only a consistent
part is added and an exception is raised
(see, e.g.,~\cite{opm} for a formal definition).
\item The ask operation is the \OzInline{\OzKeyword{case}} statement
(e.g., \OzInline{\OzKeyword{case}\OzSpace{1}Xs\OzSpace{1}\OzKeyword{of}\OzSpace{1}X|Xr\OzSpace{1}\OzKeyword{then}\OzSpace{1}...\OzSpace{1}\OzKeyword{else}\OzSpace{1}...\OzSpace{1}\OzKeyword{end}}).
It waits until the store contains enough information
to decide whether the pattern is matched (entailment)
or can never be matched (disentailment).
\end{itemize}
The above example can be written in a functional syntax.
We find that a functional syntax
often greatly improves the readability of programs.
It is especially useful when it follows the data flow,
i.e., the input and output arguments.
In Oz, the definition of \OzInline{NRev} in functional syntax
is as follows:
\begin{oz2texdisplay}\OzSpace{3}\OzKeyword{declare}\OzEol
\OzSpace{3}\OzKeyword{fun}\OzSpace{1}\OzChar\{Append\OzSpace{1}Xs\OzSpace{1}Ys\OzChar\}\OzEol
\OzSpace{6}\OzKeyword{case}\OzSpace{1}Xs\OzSpace{1}\OzKeyword{of}\OzSpace{1}nil\OzSpace{1}\OzKeyword{then}\OzSpace{1}Ys\OzEol
\OzSpace{6}[]\OzSpace{1}X|Xr\OzSpace{1}\OzKeyword{then}\OzSpace{1}X|\OzChar\{Append\OzSpace{1}Xr\OzSpace{1}Ys\OzChar\}\OzSpace{1}\OzKeyword{end}\OzEol
\OzSpace{3}\OzKeyword{end}\OzEol
\OzEol
\OzSpace{3}\OzKeyword{fun}\OzSpace{1}\OzChar\{NRev\OzSpace{1}Xs\OzChar\}\OzEol
\OzSpace{6}\OzKeyword{case}\OzSpace{1}Xs\OzSpace{1}\OzKeyword{of}\OzSpace{1}nil\OzSpace{1}\OzKeyword{then}\OzSpace{1}nil\OzEol
\OzSpace{6}[]\OzSpace{1}X|Xr\OzSpace{1}\OzKeyword{then}\OzSpace{1}\OzChar\{Append\OzSpace{1}\OzChar\{NRev\OzSpace{1}Xr\OzChar\}\OzSpace{1}[X]\OzChar\}\OzSpace{1}\OzKeyword{end}\OzEol
\OzSpace{3}\OzKeyword{end}\end{oz2texdisplay}
This is just syntactic sugar for the procedural definition.
In Oz, a function is just a shorter way of writing
a procedure where the procedure's last argument
is the function's output.
The statement \OzInline{Ys=\OzChar\{NRev\OzSpace{1}Xs\OzChar\}} has identical semantics
to the procedure call \OzInline{\OzChar\{NRev\OzSpace{1}Xs\OzSpace{1}Ys\OzChar\}}.

From the semantics outlined above,
it follows that \OzInline{Append} and \OzInline{NRev} do not search.
If there is not enough information to continue,
then the computation will simply block.
For example, take these two calls:
\begin{oz2texdisplay}\OzSpace{3}\OzKeyword{declare}\OzSpace{1}X\OzSpace{1}Y\OzSpace{1}A\OzSpace{1}B\OzSpace{1}\OzKeyword{in}\OzEol
\OzSpace{3}\OzChar\{Append\OzSpace{1}[1]\OzSpace{1}X\OzSpace{1}Y\OzChar\}\OzEol
\OzSpace{3}\OzChar\{Append\OzSpace{1}A\OzSpace{1}[2]\OzSpace{1}B\OzChar\}\end{oz2texdisplay}
(The \OzInline{\OzKeyword{declare}} ... \OzInline{\OzKeyword{in}} introduces new variables.)
The first call, \OzInline{\OzChar\{Append\OzSpace{1}[1]\OzSpace{1}X\OzSpace{1}Y\OzChar\}}, will run to completion
since \OzInline{Append} does not need the value of its second argument.
The result is the binding of \OzInline{Y} to \OzInline{1|X}.
The second call, \OzInline{\OzChar\{Append\OzSpace{1}A\OzSpace{1}[2]\OzSpace{1}B\OzChar\}}, will suspend the thread
it is executing in.
This is because the \OzInline{\OzKeyword{case}} statement does not have enough
information to decide what \OzInline{A} is.
No binding is done.
If another thread binds \OzInline{A}, then execution will continue.

This is how Oz supports deterministic logic programming.
It is purely declarative logic programming
with an operational semantics that
is fully specified and deterministic.
Programs can be translated
in a straightforward way
to a Horn clause syntax.
However, deductions are not performed by resolution.
The execution can be seen as functional programming
with logic variables and dynamic typing,
carefully designed to have a logical semantics.
Resolution was originally designed as an inference
rule for automatic theorem provers~\cite{robinson}; it is not
a necessary part of a logic programming language.

Note that there are higher-order procedures
as in a functional language,
but no higher-order logic programming,
i.e., no logic programming based on a higher-order logic.
Higher-order procedures are useful within first-order
logic programming as a tool to structure programs and build abstractions.

We find that adding logic variables
to functional programming is
an important extension for three reasons.
First, it allows to do deterministic logic programming
in a straightforward way.
Second, it increases expressiveness by
allowing powerful programming techniques based
on incomplete data structures,
such as tail-recursive append and
difference lists~\cite{difflist,artofprolog}.
The third reason is perhaps the most important:
adding concurrency to this execution model gives
a useful form of concurrent programming called
{\em declarative concurrency} (see Section~\ref{kernellang}).

\section{Nondeterministic logic programming}
\label{nondetlp}

We call {\em nondeterministic} logic programming the
situation when {\em search} is used to provide completeness.
Using search allows finding solutions
when no other algorithm is known.\footnote{To be
precise, search is a general technique that works for any problem
by giving just the problem specification,
but it can be impractical because it does brute force exploration
of a potentially large space of candidate solutions.
Search can be made more efficient by incorporating
problem-specific knowledge, e.g., games can be programmed
using alpha-beta search.}
Oz provides the \OzInline{\OzKeyword{choice}} statement
as a simple way to introduce search.
The \OzInline{\OzKeyword{choice}} statement creates
a choice point for its alternatives.

The \OzInline{\OzKeyword{choice}} statement allows
to do Prolog-style generative execution.
However, this style of programming
does not scale up to real-world search problems.\footnote{For
problems with a small search space,
they may be sufficient.
For example, a practical diagnostics generator for the VLSI-BAM
microprocessor was written in Prolog~\cite{holmerjlp,agdi}.}
In our opinion,
its primary value is pedagogical and exploratory.
That is, it can be used on small examples to explore
and understand a problem's structure.
With this understanding, a more efficient
algorithm can often be designed.
When used naively,
search will not work on large examples
due to search space explosion.

Search is a fundamental part of constraint programming.
Many techniques have been devised there
to reduce greatly the size of the search space.
Section~\ref{moresearch} gives a simple example
to illustrate some of these techniques.

Here is a nondeterministic naive reverse with \OzInline{\OzKeyword{choice}}:
\begin{oz2texdisplay}\OzSpace{3}\OzKeyword{proc}\OzSpace{1}\OzChar\{Append\OzSpace{1}Xs\OzSpace{1}Ys\OzSpace{1}Zs\OzChar\}\OzEol
\OzSpace{6}\OzKeyword{choice}\OzSpace{8}Xs=nil\OzSpace{2}Zs=Ys\OzEol
\OzSpace{6}[]\OzSpace{1}X\OzSpace{1}Xr\OzSpace{1}Zr\OzSpace{1}\OzKeyword{in}\OzSpace{1}Xs=X|Xr\OzSpace{1}Zs=X|Zr\OzSpace{1}\OzChar\{Append\OzSpace{1}Xr\OzSpace{1}Ys\OzSpace{1}Zr\OzChar\}\OzEol
\OzSpace{6}\OzKeyword{end}\OzEol
\OzSpace{3}\OzKeyword{end}\OzEol
\OzEol
\OzSpace{3}\OzKeyword{proc}\OzSpace{1}\OzChar\{NRev\OzSpace{1}Xs\OzSpace{1}Ys\OzChar\}\OzEol
\OzSpace{6}\OzKeyword{choice}\OzSpace{5}Xs=nil\OzSpace{2}Ys=nil\OzEol
\OzSpace{6}[]\OzSpace{1}X\OzSpace{1}Xr\OzSpace{1}\OzKeyword{in}\OzSpace{1}Xs=X|Xr\OzSpace{1}\OzChar\{Append\OzSpace{1}\OzChar\{NRev\OzSpace{1}Xr\OzChar\}\OzSpace{1}[X]\OzSpace{1}Ys\OzChar\}\OzEol
\OzSpace{6}\OzKeyword{end}\OzEol
\OzSpace{3}\OzKeyword{end}\end{oz2texdisplay}
(In this and all further examples, we leave out
the \OzInline{\OzKeyword{declare}} for brevity.)
Because this example does not use higher-order programming,
there is a direct translation to the Horn clause syntax of Prolog:
\begin{verbatim}
   append(Xs, Ys, Zs) :- Xs=nil, Zs=Ys.
   append(Xs, Ys, Zs) :- Xs=[X|Xr], Zs=[X|Zr], append(Xr, Ys, Zr).

   nrev(Xs, Ys) :- Xs=nil, Ys=nil.
   nrev(Xs, Ys) :- Xs=[X|Xr], nrev(Xr, Yr), append(Yr, [X], Ys).
\end{verbatim}
If the Oz program is run with depth-first search,
its semantics will be identical to the Prolog version.

\subsection*{Controlling search}

The program for nondeterministic naive reverse
can be called in many ways, e.g.,
by lazy depth-first search (similar to a Prolog top level) \footnote{Lazy
search is different from lazy evaluation in that
the program must explicitly request the next solution
(see Section~\ref{executionmodel}).},
eager search,
or interactive search (with the Explorer tool~\cite{explorer,mildversion}).
All of these search abilities are programmed
in Oz using the notion of computation space
(see Section~\ref{executionmodel}).
Often the programmer will never use
spaces directly (although he or she can), but will
use one of the many predefined search
abstractions provided in the \OzInline{Search} module
(see Section~\ref{simplesearch}).

As a first example, let us introduce an abstraction,
called {\em search object},
that is similar to a Prolog top level.
It does depth-first search and 
can be queried to obtain successive solutions.
Three steps are needed to use it: \footnote{For clarity,
we leave out syntactic short-cuts.
For example, calculating and displaying the next solution
can be written as \OzInline{\OzChar\{Browse\OzSpace{1}\OzChar\{E\OzSpace{1}next(\OzChar\$)\OzChar\}\OzChar\}}.}
\begin{oz2texdisplay}\OzSpace{3}\OzKeyword{declare}\OzSpace{1}P\OzSpace{1}E\OzSpace{1}\OzKeyword{in}\OzEol
\OzSpace{3}\OzEolComment{\OzSpace{1}1.\OzSpace{1}Define\OzSpace{1}a\OzSpace{1}new\OzSpace{1}search\OzSpace{1}query:}\OzSpace{3}\OzKeyword{proc}\OzSpace{1}\OzChar\{P\OzSpace{1}S\OzChar\}\OzSpace{1}X\OzSpace{1}Y\OzSpace{1}\OzKeyword{in}\OzSpace{1}\OzChar\{Append\OzSpace{1}X\OzSpace{1}Y\OzSpace{1}[1\OzSpace{1}2\OzSpace{1}3\OzSpace{1}4\OzSpace{1}5]\OzChar\}\OzSpace{1}S=sol(X\OzSpace{1}Y)\OzSpace{1}\OzKeyword{end}\OzEol
\OzEol
\OzSpace{3}\OzEolComment{\OzSpace{1}2.\OzSpace{1}Set\OzSpace{1}up\OzSpace{1}a\OzSpace{1}new\OzSpace{1}search\OzSpace{1}engine:}\OzSpace{3}E=\OzChar\{New\OzSpace{1}Search.object\OzSpace{1}script(P)\OzChar\}\OzEol
\OzEol
\OzSpace{3}\OzEolComment{\OzSpace{1}3.\OzSpace{1}Calculate\OzSpace{1}and\OzSpace{1}display\OzSpace{1}the\OzSpace{1}first\OzSpace{1}solution:}\OzSpace{3}\OzEolComment{\OzSpace{4}(and\OzSpace{1}others,\OzSpace{1}when\OzSpace{1}repeated)}\OzSpace{3}\OzKeyword{local}\OzSpace{1}X\OzSpace{1}\OzKeyword{in}\OzSpace{1}\OzChar\{E\OzSpace{1}next(X)\OzChar\}\OzSpace{1}\OzChar\{Browse\OzSpace{1}X\OzChar\}\OzSpace{1}\OzKeyword{end}\end{oz2texdisplay}
Let us explain each of these steps:
\begin{enumerate}
\item The procedure \OzInline{P} defines the query
and returns the solution \OzInline{S} in its single argument.
Because Oz is a higher-order language,
the query can be any statement.
In this example, the solution has two parts, \OzInline{X} and \OzInline{Y}.
We pair them together in the tuple \OzInline{sol(X\OzSpace{1}Y)}.

\item The search object is an instance of the class \OzInline{Search.object}.
The object is created with \OzInline{New} and initialized with
the message \OzInline{script(P)}.

\item The object invocation \OzInline{\OzChar\{E\OzSpace{1}next(X)\OzChar\}} finds
the next solution of the query \OzInline{P}.
If there is a solution, then \OzInline{X} is bound to a list
containing it as single element.
If there are no more solutions, then \OzInline{X} is bound to \OzInline{nil}.
\OzInline{Browse} is a tool provided by the system to display data structures.
\end{enumerate}
When running this example,
the first call
displays the solution \OzInline{[sol(nil\OzSpace{1}[1\OzSpace{1}2\OzSpace{1}3\OzSpace{1}4\OzSpace{1}5])]},
that is, a one-element list containing a solution.
Successive calls display the solutions
\OzInline{[sol([1]\OzSpace{1}[2\OzSpace{1}3\OzSpace{1}4\OzSpace{1}5])]}, ..., \OzInline{[sol([1\OzSpace{1}2\OzSpace{1}3\OzSpace{1}4\OzSpace{1}5]\OzSpace{1}nil)]}.
When there are no more solutions, then \OzInline{nil} is displayed
instead of a one-element list.

The standard Oz approach is to use search only
for problems that require it.
To solve algorithmic problems,
one does not need to learn how to use search
in the language.
This is unlike Prolog,
where search is ubiquitous:
even procedure application is defined in terms
of resolution, and thus search.
In Oz, the \OzInline{\OzKeyword{choice}} statement explicitly creates
a choice point, and search abstractions
(such as \OzInline{Search.object}, above) encapsulate and control it.
However, the \OzInline{\OzKeyword{choice}} statement by itself is a bit too simplistic,
since the choice point is statically placed.
The usual way to add choice points in Oz
is with abstractions that dynamically create
a choice point whose alternatives
depend on the state of the computation.
The heuristics used are called the {\em distribution strategy}.
For example, the procedure \OzInline{FD.distribute} allows
to specify the distribution strategy
for problems using finite domain constraints.
Section~\ref{scalable} gives an example of this approach.

\section{Concurrent logic programming}
\label{conclp}

In {\em concurrent} logic programming,
programs are written as a set of don't-care predicates
and executed concurrently.
That is, at most one clause is chosen from each predicate
invocation, in a nondeterministic way from all the clauses
whose guards are true.
This style of logic programming is incomplete,
just like deterministic logic programming.
Only a small part of the search space is explored
due to the guarded clause selection.
The advantage is that programs are concurrent,
and concurrency is essential for programs that interact with
their environment, e.g., for agents, GUI programming, OS interaction, etc.
Many algorithmic problems are of this type.
Concurrency also permits a program to be organized into parts
that execute independently and interact only when needed.
This is an important software engineering property.

In this section, we show how to do concurrent logic programming in Oz.
In fact, the full Oz language allows concurrency and search
to be used together (see Section~\ref{executionmodel}).
The clean integration of both in a single language
is one of the major strengths of Oz.
The integration was first achieved in Oz's immediate ancestor,
AKL, in 1990~\cite{akl}.
Oz shares many aspects with AKL but
improves over it in particular by being compositional
and higher-order.

\subsection{Implicit versus explicit concurrency}

In early concurrent logic programming systems,
concurrency was implicit,
driven solely by data dependencies~\cite{Shapiro:89}.
Each body goal implicitly ran in its own thread.
The hope was that this would make parallel execution easy.
But this hope has not been realized, for several reasons.
The overhead of implicit concurrency is too high,
parallelism is limited without rewriting programs,
and detecting program termination is hard.
To reduce the overhead, it is possible to do lazy thread creation,
that is, to create a new thread only when the parent thread would suspend.
This approach has a nice slogan,
``as sequential as possible, as concurrent as necessary,''
and it allows an efficient implementation.
But the approach is still inadequate
because reasoning about programs remains hard.

After implementing and experimenting with
both implicit concurrency and lazy thread creation,
the current Oz decision is to do only explicit thread creation
(see Section~\ref{historysketch}).
Explicit thread creation
simplifies debugging and reasoning about programs,
and is efficient.
Furthermore, experience shows that parallelism (i.e., speedup) is
not harder to obtain than before; it is
still the programmer's responsibility to know what parts of
the program can potentially be run in parallel.

\subsection{Concurrent producer-consumer}

A classic example of concurrent logic programming
is the asynchronous producer-consumer.
The following program asynchronously generates a stream
of integers and sums them.
A {\em stream} is a list whose tail is an unbound logic variable.
The tail can itself be bound to a stream, and so forth.
\newpage
\begin{oz2texdisplay}\OzSpace{3}\OzKeyword{proc}\OzSpace{1}\OzChar\{Generate\OzSpace{1}N\OzSpace{1}Limit\OzSpace{1}Xs\OzChar\}\OzEol
\OzSpace{6}\OzKeyword{if}\OzSpace{1}N<Limit\OzSpace{1}\OzKeyword{then}\OzSpace{1}Xr\OzSpace{1}\OzKeyword{in}\OzEol
\OzSpace{9}Xs=N|Xr\OzEol
\OzSpace{9}\OzChar\{Generate\OzSpace{1}N+1\OzSpace{1}Limit\OzSpace{1}Xr\OzChar\}\OzEol
\OzSpace{6}\OzKeyword{else}\OzSpace{1}Xs=nil\OzSpace{1}\OzKeyword{end}\OzEol
\OzSpace{3}\OzKeyword{end}\end{oz2texdisplay}
\begin{oz2texdisplay}\OzSpace{3}\OzKeyword{proc}\OzSpace{1}\OzChar\{Sum\OzSpace{1}Xs\OzSpace{1}A\OzSpace{1}S\OzChar\}\OzEol
\OzSpace{6}\OzKeyword{case}\OzSpace{1}Xs\OzEol
\OzSpace{6}\OzKeyword{of}\OzSpace{1}X|Xr\OzSpace{1}\OzKeyword{then}\OzSpace{1}\OzChar\{Sum\OzSpace{1}Xr\OzSpace{1}A+X\OzSpace{1}S\OzChar\}\OzEol
\OzSpace{6}[]\OzSpace{1}nil\OzSpace{1}\OzKeyword{then}\OzSpace{1}S=A\OzEol
\OzSpace{6}\OzKeyword{end}\OzEol
\OzSpace{3}\OzKeyword{end}\end{oz2texdisplay}
\begin{oz2texdisplay}\OzSpace{3}\OzKeyword{local}\OzSpace{1}Xs\OzSpace{1}S\OzSpace{1}\OzKeyword{in}\OzEol
\OzSpace{6}\OzKeyword{thread}\OzSpace{1}\OzChar\{Generate\OzSpace{1}0\OzSpace{1}150000\OzSpace{1}Xs\OzChar\}\OzSpace{1}\OzKeyword{end}\OzSpace{2}\OzEolComment{\OzSpace{1}Producer\OzSpace{1}thread}\OzSpace{6}\OzKeyword{thread}\OzSpace{1}\OzChar\{Sum\OzSpace{1}Xs\OzSpace{1}0\OzSpace{1}S\OzChar\}\OzSpace{1}\OzKeyword{end}\OzSpace{12}\OzEolComment{\OzSpace{1}Consumer\OzSpace{1}thread}\OzSpace{6}\OzChar\{Browse\OzSpace{1}S\OzChar\}\OzEol
\OzSpace{3}\OzKeyword{end}\end{oz2texdisplay}
This executes as expected in the concurrent logic programming framework.
The producer, \OzInline{Generate},
and the consumer, \OzInline{Sum}, run in their own threads.
They communicate through the shared variable \OzInline{Xs},
which is a stream of integers.
The \OzInline{\OzKeyword{case}} statement in \OzInline{Sum} synchronizes on \OzInline{Xs} being bound to a value.

This example has exactly one producer feeding exactly one consumer.
It therefore does not need a nondeterministic choice.
More general cases do, e.g., a client-server application
with more than one client feeding a server.
Without additional information,
the server never knows which client will send the next request.
Nondeterministic choice can be added directly to the language,
e.g., the \OzInline{WaitTwo} operation of Section~\ref{kernellang}.
It turns out to be more practical to add state instead.
Then nondeterministic choice is a consequence of having
both state and concurrency, as explained in Section~\ref{state}.

\subsection{Lazy producer-consumer}
\label{lazyexample}

In the above producer-consumer example,
it is the producer that decides how many
list elements to generate.
This is called {\em supply-driven} or {\em eager} execution.
This is an efficient technique if the total amount of work
is finite and
does not use many system resources (e.g., memory or calculation time).
On the other hand, if the total work potentially uses many
resources, then it may be better to use {\em demand-driven}
or {\em lazy} execution.
With lazy execution, the consumer decides
how many list elements to generate.
If an extremely large or a potentially unbounded number
of list elements are needed, then lazy execution will
use many fewer system resources at any given point in time.
Problems that are impractical with eager execution can become
practical with lazy execution.

Lazy execution can be implemented in two ways in Oz.
The first way, which is applicable to any language,
is to use {\em explicit triggers}.
The producer and consumer are modified so that
the consumer asks the producer for additional list elements.
In our example,
the simplest way is to use logic variables as explicit triggers.
The consumer binds the end of a stream to \OzInline{X|\OzChar\_}.
The producer waits for this and binds \OzInline{X} to the next
list element.

Explicit triggers are cumbersome because they require
the producer to accept explicit communications from the consumer.
A better way is for the language to support laziness directly.
That is, the language semantics would ensure that a function is
evaluated only if its result were needed.
Oz supports this syntactically
by annotating the function as ``\OzInline{lazy}''.
Here is how to do the previous
example with a lazy function that generates a potentially infinite list:
\begin{oz2texdisplay}\OzSpace{3}\OzKeyword{fun}\OzSpace{1}lazy\OzSpace{1}\OzChar\{Generate\OzSpace{1}N\OzChar\}\OzEol
\OzSpace{6}N|\OzChar\{Generate\OzSpace{1}N+1\OzChar\}\OzEol
\OzSpace{3}\OzKeyword{end}\end{oz2texdisplay}
\begin{oz2texdisplay}\OzSpace{3}\OzKeyword{proc}\OzSpace{1}\OzChar\{Sum\OzSpace{1}Xs\OzSpace{1}Limit\OzSpace{1}A\OzSpace{1}S\OzChar\}\OzEol
\OzSpace{6}\OzKeyword{if}\OzSpace{1}Limit>0\OzSpace{1}\OzKeyword{then}\OzEol
\OzSpace{9}\OzKeyword{case}\OzSpace{1}Xs\OzEol
\OzSpace{9}\OzKeyword{of}\OzSpace{1}X|Xr\OzSpace{1}\OzKeyword{then}\OzEol
\OzSpace{12}\OzChar\{Sum\OzSpace{1}Xr\OzSpace{1}Limit-1\OzSpace{1}A+X\OzSpace{1}S\OzChar\}\OzEol
\OzSpace{9}\OzKeyword{end}\OzEol
\OzSpace{6}\OzKeyword{else}\OzSpace{1}S=A\OzSpace{1}\OzKeyword{end}\OzEol
\OzSpace{3}\OzKeyword{end}\end{oz2texdisplay}
\begin{oz2texdisplay}\OzSpace{3}\OzKeyword{local}\OzSpace{1}Xs\OzSpace{1}S\OzSpace{1}\OzKeyword{in}\OzEol
\OzSpace{6}\OzKeyword{thread}\OzSpace{1}Xs=\OzChar\{Generate\OzSpace{1}0\OzChar\}\OzSpace{1}\OzKeyword{end}\OzEol
\OzSpace{6}\OzKeyword{thread}\OzSpace{1}\OzChar\{Sum\OzSpace{1}Xs\OzSpace{1}150000\OzSpace{1}0\OzSpace{1}S\OzChar\}\OzSpace{1}\OzKeyword{end}\OzEol
\OzSpace{6}\OzChar\{Browse\OzSpace{1}S\OzChar\}\OzEol
\OzSpace{3}\OzKeyword{end}\end{oz2texdisplay}
Here the consumer, \OzInline{Sum}, decides
how many list elements should be generated.
The addition \OzInline{A+X} implicitly triggers
the generation of a new list element \OzInline{X}.
Lazy execution is part of the Oz execution
model; Section~\ref{lazykernel} explains how it works.

\subsection{Coroutining}

Sequential systems often support coroutining
as a simple way to get some of the abilities of concurrency.
Coroutining is a form of non-preemptive concurrency
in which a single locus of control is switched manually
between different parts of a program.
In our experience, a system with efficient
preemptive concurrency almost never needs coroutining.

Most modern Prolog systems support coroutining.
The coroutining is either supported directly,
as in IC-Prolog~\cite{CM79:,Clar82a},
or indirectly
by means of an operation called {\tt freeze}
which provides data-driven computation.
The {\tt freeze(X,G)} operation,
sometimes called {\tt geler(X,G)}
from Prolog II which pioneered it~\cite{colmerauer82},
sets up the system to invoke the goal {\tt G} when
the variable {\tt X} is bound~\cite{artofprolog}.
With {\tt freeze} it is possible to have ``non-preemptive threads'' that
explicitly hand over control to each other by binding variables.
Because Prolog's search is based on global backtracking,
the ``threads'' are not independent: if a thread backtracks,
then other threads may be forced to backtrack as well.
Prolog programming techniques that depend on backtracking,
such as search, deep conditionals, and exceptions,
cannot be used if the program has to switch between threads.

\section{Explicit state}
\label{state}

From a theoretical point of view,
explicit state has often been considered
a forbidden fruit in logic programming.
We find that using explicit state
is important for fundamental reasons
related to program modularity
(see Chapter 4 of~\cite{book}).

There exist tools to use state in Prolog while keeping
a logical semantics when possible.
See for example SICStus Objects~\cite{sicstus}, Prolog++~\cite{moss},
and the Logical State Threads package~\cite{lst97}.
An ancestor of the latter was used to help write
the Aquarius Prolog compiler~\cite{edcg,aquarius}.

Functional programmers have also incorporated
state into functional languages, e.g., by means of
\texttt{set} operations in LISP/Scheme~\cite{commonlisp,abelson2},
references in ML~\cite{MilnerTofteHarper:90},
and monads in Haskell~\cite{Wadler:92c}.

\subsection{Cells (mutable references)}

State is an explicit part of the basic execution model in Oz.
The model defines the concept of {\em cell},
which is a kind of mutable reference.
A cell is a pair of a name \OzInline{C} and a reference \OzInline{X}.
There are two operations on cells:
\begin{oz2texdisplay}\OzSpace{2}\OzChar\{NewCell\OzSpace{1}X\OzSpace{1}C\OzChar\}\OzSpace{4}\OzEolComment{\OzSpace{1}Create\OzSpace{1}new\OzSpace{1}cell\OzSpace{1}with\OzSpace{1}name\OzSpace{1}C\OzSpace{1}and\OzSpace{1}content\OzSpace{1}X}\OzSpace{2}\OzChar\{Exchange\OzSpace{1}C\OzSpace{1}X\OzSpace{1}Y\OzChar\}\OzSpace{1}\OzEolComment{\OzSpace{1}Update\OzSpace{1}content\OzSpace{1}to\OzSpace{1}Y\OzSpace{1}and\OzSpace{1}bind\OzSpace{1}X\OzSpace{1}to\OzSpace{1}old\OzSpace{1}content}\end{oz2texdisplay}
Each \OzInline{Exchange} atomically accesses the current content and
defines a new content.

Oz has a full-featured concurrent object system
which is completely defined in terms
of cells~\cite{HenzDiss:97,HenzOFCCP:97}.
The object system includes multiple inheritance,
fine-grained method access control,
and first-class messages.
Sections~\ref{executionmodel} gives more information
about cells and explains how they underlie the object system.

\subsection{Ports (communication channels)}

In this section we present another,
equivalent way to add state to the basic model.
This is the concept of {\em port}, which was pioneered by AKL.
A port is a pair of a name \OzInline{P} and a stream \OzInline{Xs}~\cite{portpaper}.
There are two operations on ports:
\begin{oz2texdisplay}\OzSpace{3}\OzChar\{NewPort\OzSpace{1}Xs\OzSpace{1}P\OzChar\}\OzSpace{2}\OzEolComment{\OzSpace{1}Create\OzSpace{1}new\OzSpace{1}port\OzSpace{1}with\OzSpace{1}name\OzSpace{1}P\OzSpace{1}and\OzSpace{1}stream\OzSpace{1}Xs}\OzSpace{3}\OzChar\{Send\OzSpace{1}P\OzSpace{1}X\OzChar\}\OzSpace{6}\OzEolComment{\OzSpace{1}Add\OzSpace{1}X\OzSpace{1}to\OzSpace{1}port's\OzSpace{1}stream\OzSpace{1}asynchronously}\end{oz2texdisplay}
Each \OzInline{Send} asynchronously adds one more element to the port's stream.
The port keeps an internal reference to the stream's unbound tail.
Repeated sends in the same thread cause the elements to appear
in the same order as the sends.
There are no other ordering constraints on the stream.

Using ports gives us the ability to have named active objects.
An {\em active object}, in its simplest form,
pairs an object with a thread.
The thread reads a stream of internal and external messages,
and invokes the object for each message.
The Erlang language is based on this idea~\cite{erlang}.
Erlang extends it by adding
to each object a mailbox that does retrieval by pattern matching.

With cells it is natural to define
non-active objects, called {\em passive objects},
shared between threads.
With ports it is natural to define
active objects that send messages to each other.
From a theoretical point of view,
these two programming styles have the same expressiveness,
since cells and ports can be defined in terms of each other
without changing time or space complexity~\cite{HenzOFCCP:97,needhamlauer78}.
They differ in practice,
since depending on the application
one style might be more convenient than the other.
Database applications, which are centered around a shared data repository,
find the shared object style natural.
Multi-agent applications,
defined in terms of collaborating active entities,
find the active object style natural.


\subsection{Relevance to concurrent logic programming}

From the perspective of concurrent logic programming,
explicit state amounts to the addition of
a constant-time $n$-way stream merge,
where $n$ can grow arbitrarily large at run-time.
That is, any number
of threads can concurrently send to the same port,
and each send will take constant time.
This can be seen as the ability to give
an {\em identity} to an active object.
The identity is a first-class value:
it can be stored in a data structure
and can be passed as an argument.
It is enough to know the identity to send
a message to the active object.

Without explicit state it impossible
to build this kind of merge.
If $n$ is known only at run-time,
the only solution is to build a tree of stream mergers.
With $n$ senders, this multiplies
the message sending time by $O(\log n)$.
We know of no simple way to solve this problem
other than by adding explicit state to the execution model.

\subsection{Creating an active object}

Here is an example that uses a port to make an active object:
\begin{oz2texdisplay}\OzSpace{3}\OzKeyword{proc}\OzSpace{1}\OzChar\{DisplayStream\OzSpace{1}Xs\OzChar\}\OzEol
\OzSpace{6}\OzKeyword{case}\OzSpace{1}Xs\OzSpace{1}\OzKeyword{of}\OzSpace{1}X|Xr\OzSpace{1}\OzKeyword{then}\OzSpace{1}\OzChar\{Browse\OzSpace{1}X\OzChar\}\OzSpace{1}\OzChar\{DisplayStream\OzSpace{1}Xr\OzChar\}\OzEol
\OzSpace{6}\OzKeyword{else}\OzSpace{1}\OzKeyword{skip}\OzSpace{1}\OzKeyword{end}\OzEol
\OzSpace{3}\OzKeyword{end}\OzEol
\OzEol
\OzSpace{3}\OzKeyword{declare}\OzSpace{1}P\OzSpace{1}\OzKeyword{in}\OzSpace{4}\OzEolComment{\OzSpace{1}P\OzSpace{1}has\OzSpace{1}global\OzSpace{1}scope}\OzSpace{3}\OzKeyword{local}\OzSpace{1}Xs\OzSpace{1}\OzKeyword{in}\OzSpace{5}\OzEolComment{\OzSpace{1}Xs\OzSpace{1}has\OzSpace{1}local\OzSpace{1}scope}\OzSpace{6}\OzChar\{NewPort\OzSpace{1}Xs\OzSpace{1}P\OzChar\}\OzEol
\OzSpace{6}\OzKeyword{thread}\OzSpace{1}\OzChar\{DisplayStream\OzSpace{1}Xs\OzChar\}\OzSpace{1}\OzKeyword{end}\OzEol
\OzSpace{3}\OzKeyword{end}\end{oz2texdisplay}
Sending to \OzInline{P} sends to the active object.
Any number of clients can send to the active object
concurrently:
\newpage
\begin{oz2texdisplay}\OzSpace{3}\OzKeyword{thread}\OzSpace{1}\OzChar\{Send\OzSpace{1}P\OzSpace{1}1\OzChar\}\OzSpace{1}\OzChar\{Send\OzSpace{1}P\OzSpace{1}2\OzChar\}\OzSpace{1}...\OzSpace{1}\OzKeyword{end}\OzSpace{3}\OzEolComment{\OzSpace{1}Client\OzSpace{1}1}\OzSpace{3}\OzKeyword{thread}\OzSpace{1}\OzChar\{Send\OzSpace{1}P\OzSpace{1}a\OzChar\}\OzSpace{1}\OzChar\{Send\OzSpace{1}P\OzSpace{1}b\OzChar\}\OzSpace{1}...\OzSpace{1}\OzKeyword{end}\OzSpace{3}\OzEolComment{\OzSpace{1}Client\OzSpace{1}2}\end{oz2texdisplay}
The elements \OzInline{1}, \OzInline{2}, \OzInline{a}, \OzInline{b}, etc., will appear
fairly on the stream \OzInline{Xs}.
Port fairness is guaranteed because of thread fairness
in the Mozart implementation.


Here is a more compact way to define the active object's thread:
\begin{oz2texdisplay}\OzSpace{3}\OzKeyword{thread}\OzEol
\OzSpace{6}\OzChar\{ForAll\OzSpace{1}Xs\OzSpace{1}\OzKeyword{proc}\OzSpace{1}\OzChar\{\OzChar\$\OzSpace{1}X\OzChar\}\OzSpace{1}\OzChar\{Browse\OzSpace{1}X\OzChar\}\OzSpace{1}\OzKeyword{end}\OzChar\}\OzEol
\OzSpace{3}\OzKeyword{end}\end{oz2texdisplay}
The notation \OzInline{\OzKeyword{proc}\OzSpace{1}\OzChar\{\OzChar\$\OzSpace{1}X\OzChar\}\OzSpace{1}...\OzSpace{1}\OzKeyword{end}} defines
an {\em anonymous} procedure value,
which is not bound to any identifier.
\OzInline{ForAll} is a higher-order procedure
that applies a unary procedure
to all elements of a list.
\OzInline{ForAll} keeps the dataflow synchronization
when traversing the list.
This is an example how higher-orderness can be used
to modularize a program: the iteration
is separated from the action to be performed
on each iteration.

\section{More on search}
\label{moresearch}

We have already introduced search
in Section~\ref{nondetlp}
by means of the \OzInline{\OzKeyword{choice}} statement and the
lazy depth-first abstraction \OzInline{Search.object}.
The programming style shown there is too limited
for many realistic problems.
This section shows
how to make search more practical in Oz.
We only scratch the surface of
how to use search in Oz; for more information
we suggest the Finite Domain and Finite Set
tutorials in the Mozart system documentation~\cite{fdtutorial,fstutorial}.

\subsection{Aggregate search}
\label{aggreg}

One of the powerful features of Prolog
is its ability to generate aggregates based
on complex queries, through the
built-in operations {\tt setof/3} and {\tt bagof/3}.
These are easy to do in Oz; they are
just special cases of search abstractions.
In this section
we show how to implement {\tt bagof/3}.
Consider the following small biblical database
(taken from~\cite{artofprolog}):
\begin{oz2texdisplay}\OzSpace{3}\OzKeyword{proc}\OzSpace{1}\OzChar\{Father\OzSpace{1}F\OzSpace{1}C\OzChar\}\OzEol
\OzSpace{6}\OzKeyword{choice}\OzSpace{1}F=terach\OzSpace{2}C=abraham\OzEol
\OzSpace{10}[]\OzSpace{1}F=terach\OzSpace{2}C=nachor\OzEol
\OzSpace{10}[]\OzSpace{1}F=terach\OzSpace{2}C=haran\OzEol
\OzSpace{10}[]\OzSpace{1}F=abraham\OzSpace{1}C=isaac\OzEol
\OzSpace{10}[]\OzSpace{1}F=haran\OzSpace{3}C=lot\OzEol
\OzSpace{10}[]\OzSpace{1}F=haran\OzSpace{3}C=milcah\OzEol
\OzSpace{10}[]\OzSpace{1}F=haran\OzSpace{3}C=yiscah\OzEol
\OzSpace{6}\OzKeyword{end}\OzEol
\OzSpace{3}\OzKeyword{end}\end{oz2texdisplay}
Now consider the following Prolog predicate:
\begin{verbatim}
   children1(X, Kids) :- bagof(K, father(X,K), Kids).
\end{verbatim}
This is defined in Oz as follows:
\begin{oz2texdisplay}\OzSpace{3}\OzKeyword{proc}\OzSpace{1}\OzChar\{ChildrenFun\OzSpace{1}X\OzSpace{1}Kids\OzChar\}\OzEol
\OzSpace{3}F\OzSpace{1}\OzKeyword{in}\OzEol
\OzSpace{6}\OzKeyword{proc}\OzSpace{1}\OzChar\{F\OzSpace{1}K\OzChar\}\OzSpace{1}\OzChar\{Father\OzSpace{1}X\OzSpace{1}K\OzChar\}\OzSpace{1}\OzKeyword{end}\OzEol
\OzSpace{6}\OzChar\{Search.base.all\OzSpace{1}F\OzSpace{1}Kids\OzChar\}\OzEol
\OzSpace{3}\OzKeyword{end}\end{oz2texdisplay}
The procedure \OzInline{F} is a lexically-scoped closure:
it has the external reference \OzInline{X} hidden inside.
This can be written more compactly
with an anonymous procedure value:
\begin{oz2texdisplay}\OzSpace{3}\OzKeyword{proc}\OzSpace{1}\OzChar\{ChildrenFun\OzSpace{1}X\OzSpace{1}Kids\OzChar\}\OzEol
\OzSpace{6}\OzChar\{Search.base.all\OzSpace{1}\OzKeyword{proc}\OzSpace{1}\OzChar\{\OzChar\$\OzSpace{1}K\OzChar\}\OzSpace{1}\OzChar\{Father\OzSpace{1}X\OzSpace{1}K\OzChar\}\OzSpace{1}\OzKeyword{end}\OzSpace{1}Kids\OzChar\}\OzEol
\OzSpace{3}\OzKeyword{end}\end{oz2texdisplay}
The \OzInline{Search.base.all} abstraction takes a one-argument
procedure and returns the list of all solutions to the procedure.
The example call:
\begin{oz2texdisplay}\OzSpace{3}\OzChar\{Browse\OzSpace{1}\OzChar\{ChildrenFun\OzSpace{1}terach\OzChar\}\OzChar\}\end{oz2texdisplay}
returns \OzInline{[abraham\OzSpace{1}nachor\OzSpace{1}haran]}.
The \OzInline{ChildrenFun} definition is deterministic;
if called with a known \OzInline{X} then it returns \OzInline{Kids}.
To search over different values of \OzInline{X}
we give the following definition instead:
\begin{oz2texdisplay}\OzSpace{3}\OzKeyword{proc}\OzSpace{1}\OzChar\{ChildrenRel\OzSpace{1}X\OzSpace{1}Kids\OzChar\}\OzEol
\OzSpace{6}\OzChar\{Father\OzSpace{1}X\OzSpace{1}\OzChar\_\OzChar\}\OzEol
\OzSpace{6}\OzChar\{Search.base.all\OzSpace{1}\OzKeyword{proc}\OzSpace{1}\OzChar\{\OzChar\$\OzSpace{1}K\OzChar\}\OzSpace{1}\OzChar\{Father\OzSpace{1}X\OzSpace{1}K\OzChar\}\OzSpace{1}\OzKeyword{end}\OzSpace{1}Kids\OzChar\}\OzEol
\OzSpace{3}\OzKeyword{end}\end{oz2texdisplay}
The call \OzInline{\OzChar\{Father\OzSpace{1}X\OzSpace{1}\OzChar\_\OzChar\}} creates a choice point on \OzInline{X}.
The ``\OzInline{\OzChar\_}'' is syntactic sugar for \OzInline{\OzKeyword{local}\OzSpace{1}X\OzSpace{1}\OzKeyword{in}\OzSpace{1}X\OzSpace{1}\OzKeyword{end}},
which is just a new variable with a tiny scope.
The example call:
\begin{oz2texdisplay}\OzSpace{3}\OzChar\{Browse\OzSpace{1}\OzChar\{Search.base.all\OzEol
\OzSpace{5}\OzKeyword{proc}\OzSpace{1}\OzChar\{\OzChar\$\OzSpace{1}Q\OzChar\}\OzSpace{1}X\OzSpace{1}Kids\OzSpace{1}\OzKeyword{in}\OzSpace{1}\OzChar\{ChildrenRel\OzSpace{1}X\OzSpace{1}Kids\OzChar\}\OzSpace{1}Q=sol(X\OzSpace{1}Kids)\OzSpace{1}\OzKeyword{end}\OzChar\}\OzChar\}\end{oz2texdisplay}
returns:
\begin{oz2texdisplay}\OzSpace{3}[sol(terach\OzSpace{1}[abraham\OzSpace{1}nachor\OzSpace{1}haran])\OzEol
\OzSpace{4}sol(terach\OzSpace{1}[abraham\OzSpace{1}nachor\OzSpace{1}haran])\OzEol
\OzSpace{4}sol(terach\OzSpace{1}[abraham\OzSpace{1}nachor\OzSpace{1}haran])\OzEol
\OzSpace{4}sol(abraham\OzSpace{1}[isaac])\OzEol
\OzSpace{4}sol(haran\OzSpace{1}[lot\OzSpace{1}milcah\OzSpace{1}yiscah])\OzEol
\OzSpace{4}sol(haran\OzSpace{1}[lot\OzSpace{1}milcah\OzSpace{1}yiscah])\OzEol
\OzSpace{4}sol(haran\OzSpace{1}[lot\OzSpace{1}milcah\OzSpace{1}yiscah])]\end{oz2texdisplay}
In Prolog, {\tt bagof} can use existential quantification.
For example, the Prolog predicate:
\begin{verbatim}
   children2(Kids) :- bagof(K, X^father(X,K), Kids).
\end{verbatim}
collects all children such that there exists a father.
This is defined in Oz as follows:
\begin{oz2texdisplay}\OzSpace{3}\OzKeyword{proc}\OzSpace{1}\OzChar\{Children2\OzSpace{1}Kids\OzChar\}\OzEol
\OzSpace{6}\OzChar\{Search.base.all\OzSpace{1}\OzKeyword{proc}\OzSpace{1}\OzChar\{\OzChar\$\OzSpace{1}K\OzChar\}\OzSpace{1}\OzChar\{Father\OzSpace{1}\OzChar\_\OzSpace{1}K\OzChar\}\OzSpace{1}\OzKeyword{end}\OzSpace{1}Kids\OzChar\}\OzEol
\OzSpace{3}\OzKeyword{end}\end{oz2texdisplay}
The Oz solution uses \OzInline{\OzChar\_} to add a new existentially-scoped variable.
The Prolog solution, on the other hand, introduces
a new concept, namely the ``existential quantifier'' notation \verb+X^+,
which only has meaning in terms of {\tt setof/3} and {\tt bagof/3}.
The fact that this notation denotes
an existential quantifier is arbitrary.
The Oz solution introduces no new concepts.
It really does existential quantification inside the search query.

\subsection{Simple search procedures}
\label{simplesearch}

The procedure \OzInline{Search.base.all} shown in the previous section
is just one of a whole set of search procedures provided by Oz
for elementary nondeterministic logic programming.
We give a short overview; for more information
see the System Modules documentation in the
Mozart system~\cite{systemmodules}.
All procedures take as argument a unary procedure \OzInline{\OzChar\{P\OzSpace{1}X\OzChar\}},
where \OzInline{X} is bound to a solution.
Except for lazy search,
they all provide depth-first search (one and all solution)
and branch-and-bound search (with a cost function).
Here are the procedures:
\begin{itemize}
\item {\bf Basic search}.  This is the simplest to use;
no extra parameters are needed.

\item {\bf General-purpose search}.  This allows parameterizing the search
with the maximal recomputation distance
(for optimizing time and memory use),
with an asynchronous kill procedure to allow stopping infinite searches,
and with the option to return solutions either directly or
encapsulated in computation spaces (see Section~\ref{compspaces}).
Search implemented with spaces 
using strategies combining cloning and recomputation
is competitive in time and memory
with systems using trailing~\cite{Schulte:ICLP99}.
Using encapsulation, 
general-purpose search can be used
as a primitive to build more sophisticated searches.

\item {\bf Parallel search}.
When provided with a list of machines, this will spread out the
search process over these machines transparently.
We have benchmarked realistic constraint problems
on up to six machines with linear speedups~\cite{Schulte:02,SchulteDiss:00,Schulte:00b}.
The order in which the search tree is explored is nondeterministic,
and is likely to be different from depth-first or breadth-first.
If the entire tree is explored, then the number of exploration steps
is the same as depth-first search.
The speedup is a consequence of this fact 
together with the spreading of work.

\item {\bf Lazy search}.  This provides next solution and
last solution operations,
a stop operation, and a close operation.
This is a first-class Prolog top level.

\item {\bf Explorer search}.
The Explorer is a concurrent graphic tool that allows to
visualize and interactively guide
the search process~\cite{explorer,Schulte:97}.
It is invaluable for search debugging and for gaining understanding of
the structure of the problem.
\end{itemize}
All of these procedures are implemented in Oz using
computation spaces (see Section~\ref{compspaces}).
Many more specialized search procedures are available for constraint
programming, and the user can easily define his or her own.

\begin{figure}
\figrule
\begin{oz2texdisplay}\OzSpace{3}\OzKeyword{functor}\OzSpace{1}Fractions\OzSpace{3}\OzEolComment{\OzSpace{1}Name\OzSpace{1}of\OzSpace{1}module\OzSpace{1}specification}\OzSpace{6}\OzKeyword{import}\OzSpace{1}FD\OzSpace{8}\OzEolComment{\OzSpace{1}Needs\OzSpace{1}the\OzSpace{1}module\OzSpace{1}FD}\OzSpace{6}\OzKeyword{export}\OzSpace{1}script:P\OzSpace{2}\OzEolComment{\OzSpace{1}Procedure\OzSpace{1}P\OzSpace{1}defines\OzSpace{1}the\OzSpace{1}problem}\OzSpace{3}\OzKeyword{define}\OzEol
\OzSpace{6}\OzKeyword{proc}\OzSpace{1}\OzChar\{P\OzSpace{1}Sol\OzChar\}\OzEol
\OzSpace{9}A\OzSpace{1}B\OzSpace{1}C\OzSpace{1}D\OzSpace{1}E\OzSpace{1}F\OzSpace{1}G\OzSpace{1}H\OzSpace{1}I\OzSpace{1}BC\OzSpace{1}EF\OzSpace{1}HI\OzEol
\OzSpace{6}\OzKeyword{in}\OzEol
\OzSpace{9}Sol=sol(a:A\OzSpace{1}b:B\OzSpace{1}c:C\OzSpace{1}d:D\OzSpace{1}e:E\OzSpace{1}f:F\OzSpace{1}g:G\OzSpace{1}h:H\OzSpace{1}i:I)\OzEol
\OzSpace{9}BC=\OzChar\{FD.decl\OzChar\}\OzSpace{1}EF=\OzChar\{FD.decl\OzChar\}\OzSpace{1}HI=\OzChar\{FD.decl\OzChar\}\OzEol
\OzSpace{9}\OzEolComment{\OzChar\%\OzChar\%\OzSpace{1}The\OzSpace{1}constraints:}\OzSpace{9}Sol:::1\OzChar\#9\OzSpace{10}\OzEolComment{\OzSpace{1}Each\OzSpace{1}letter\OzSpace{1}represents\OzSpace{1}a\OzSpace{1}digit}\OzSpace{9}\OzChar\{FD.distinct\OzSpace{1}Sol\OzChar\}\OzSpace{2}\OzEolComment{\OzSpace{1}All\OzSpace{1}digits\OzSpace{1}are\OzSpace{1}different}\OzSpace{9}BC=:10*B+C\OzSpace{9}\OzEolComment{\OzSpace{1}Definition\OzSpace{1}of\OzSpace{1}BC}\OzSpace{9}EF=:10*E+F\OzSpace{9}\OzEolComment{\OzSpace{1}Definition\OzSpace{1}of\OzSpace{1}EF}\OzSpace{9}HI=:10*H+I\OzSpace{9}\OzEolComment{\OzSpace{1}Definition\OzSpace{1}of\OzSpace{1}HI}\OzSpace{9}A*EF*HI+D*BC*HI+G*BC*EF=:BC*EF*HI\OzSpace{2}\OzEolComment{\OzSpace{1}Main\OzSpace{1}constraint}\OzSpace{9}\OzEolComment{\OzChar\%\OzChar\%\OzSpace{1}The\OzSpace{1}distribution\OzSpace{1}strategy:}\OzSpace{9}\OzChar\{FD.distribute\OzSpace{1}ff\OzSpace{1}Sol\OzChar\}\OzEol
\OzSpace{6}\OzKeyword{end}\OzEol
\OzSpace{3}\OzKeyword{end}\end{oz2texdisplay}
\caption{A more scalable way to do search}
\label{scalablesearch}
\figrule
\end{figure}

\subsection{A more scalable way to do search}
\label{scalable}

The original motivation for doing search in Oz comes
from constraint programming.
To do search, Oz uses a concurrent version of
the following approach, which is commonly used
in (sequential) constraint logic programming:
\begin{itemize}
\item First, declaratively specify the problem by means of constraints.
The constraints have an operational as well as a declarative reading.
The operational reading specifies the deductions that the constraints
can make locally.
To get good results, the constraints must be able
to do deductions over big parts of the problem
(i.e., deductions that consider many problem variables together).
Such constraints are called ``global''.
\item Second, define and explore the search tree in a controlled way,
using heuristics to exploit the problem structure.
The general technique is called ``propagate and distribute'',
because it alternates propagation steps
(where the constraints propagate information amongst themselves)
with distribution steps
(where a choice is selected in a choice point).\footnote{The term
``distribution'' as used here refers to the distribution of $\wedge$ over $\vee$
in the logical formula $c \wedge (a \vee b)$
and has nothing to do with distributed systems consisting of
independent computers connected by a network.}
See, e.g.,~\cite{constraints96}, for more explanation.
\end{itemize}
This approach is widely applicable.
For example, it is being applied successfully
to computational linguistics~\cite{duchier-mol6,CP-NL,DucGarNie99}.
In this section, we show how to solve a simple integer puzzle.
Consider the problem of finding nine distinct digits $A$, $B$, ..., $I$,
so that the following equation holds:
\begin{quote}
$A / BC + D / EF + G / HI = 1$
\end{quote}
Here, $BC$ represents the integer $10 \times B+C$.
Figure~\ref{scalablesearch} shows how to specify this
as a constraint problem.
The unary procedure \OzInline{\OzChar\{P\OzSpace{1}Sol\OzChar\}} fully defines the problem
and the distribution strategy.
The problem is specified as a conjunction of
constraints on \OzInline{Sol},
which is bound to a record that contains the solution.\footnote{To be
precise, \OzInline{Sol} is bound to a feature tree,
which is a logical formulation of a record.}
The record has fields \OzInline{a}, ..., \OzInline{i},
one for each solution variable.
The problem constraints
are expressed in terms of {\em finite domains},
i.e., finite sets of integers.
For example, the notation \OzInline{1\OzChar\#9} represents the set $\{1, 2, ..., 9\}$.
The constraints are defined
in the module \OzInline{FD}~\cite{systemmodules}.
For example, \OzInline{FD.distinct} is a global constraint that asserts
that all its component variables are distinct integers.

\OzInline{Fractions} defines \OzInline{P} inside a {\em functor},
i.e., a module specification, in Oz terminology.
The functor defines explicitly what
process-specific resources the module needs.
This allows us to set up a parallel search engine
that spreads the constraint solving over several
machines~\cite{modules-98}.
If execution is always in the same process,
then the functor is not needed and it is enough to
define the procedure \OzInline{P}.
Let's set up a parallel search engine:
\begin{oz2texdisplay}\OzSpace{3}E=\OzChar\{New\OzSpace{1}Search.parallel\OzEol
\OzSpace{10}init(adventure:1\OzChar\#rsh\OzSpace{1}galley:1\OzChar\#rsh\OzSpace{1}norge:1\OzChar\#rsh)\OzChar\}\end{oz2texdisplay}
This sets up an engine on the three machines
\OzInline{adventure}, \OzInline{galley}, and \OzInline{norge}.
The engine is implemented using computation spaces
(see Section~\ref{compspaces})
and Mozart's support for distributed computing (see~\cite{ngc98}).
A single process is created on each of these machines
using the remote shell operation \OzInline{rsh}
(other operations are possible including secure shell \OzInline{ssh}
for secure communication and local shell \OzInline{sh}
for shared-memory multiprocessors).
The following command does parallel search
on the problem specified in \OzInline{Fractions}:
\begin{oz2texdisplay}\OzSpace{3}\OzKeyword{local}\OzSpace{1}X\OzSpace{1}\OzKeyword{in}\OzSpace{1}\OzChar\{E\OzSpace{1}all(Fractions\OzSpace{1}X)\OzChar\}\OzSpace{1}\OzChar\{Browse\OzSpace{1}X\OzChar\}\OzSpace{1}\OzKeyword{end}\end{oz2texdisplay}
This installs the functor \OzInline{Fractions} on each of
the three machines and generates all the solutions.
This is an example of a more scalable way to do search:
first use global constraints and search heuristics,
and then use parallel execution if necessary for performance.

Oz is currently one of the
most advanced languages for programming search.
Competitors are
CLAIRE and SaLSA~\cite{claire,salsa,caseau_meta_cp99}
and OPL~\cite{opl}.
Search is also an important part
of constraint programming in general~\cite{constraintprogramming}.

%

\section{The Oz execution model}
\label{executionmodel}

So far, we have highlighted different parts of Oz without
showing how they interact,
something like the proverbial elephant that
is different things to different people.
This section gives the simple execution model that underlies it all.
We define the execution model in
terms of a {\em store} (Section~\ref{store})
and a {\em kernel language} (Section~\ref{kernellang}).
Section~\ref{multiparadigm} explains how
different subsets of the kernel language
support different programming paradigms.
The section also explains why supporting
multiple paradigms is useful.
Finally, Section~\ref{compspaces} defines
computation spaces and how they are used
to program search.

\begin{figure}
\figrule
\psfig{file=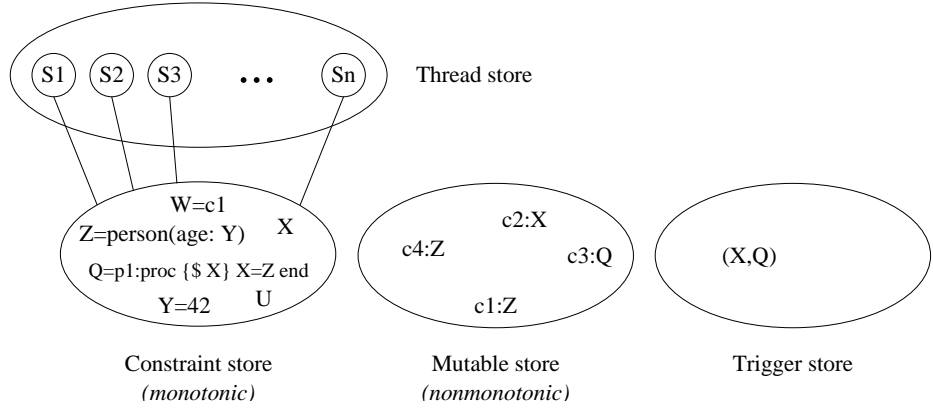,width=\textwidth}
\caption{The Oz store.}\label{storefig}
\figrule
\end{figure}

\subsection{The store}
\label{store}

The Oz store
consists of four parts (see Figure~\ref{storefig}):
a thread store, a constraint store,
a mutable store, and a trigger store.
The constraint store contains
equality constraints over the domain of rational trees.
In other words, this store contains
logic variables that are either unbound or bound.
A bound variable
references a term (i.e., atom, record, procedure, or name)
whose arguments themselves may be bound or unbound.
Unbound variables can be bound to unbound variables,
in which case they become identical references.
The constraint store is {\em monotonic},
i.e., bindings can only be added, not removed or changed.

The mutable store consists of
mutable references into the constraint store.
Mutable references 
are also called {\em cells}~\cite{HenzOFCCP:97}.
A mutable reference consists of two parts:
its {\em name}, which is a value,
and its {\em content}, which is
a reference into the constraint store.
The mutable store is {\em nonmonotonic}
because a mutable reference can be changed.

The trigger store consists of {\em triggers},
which are pairs of variables and one-argument procedures.
Since these triggers are part of the basic execution model,
they are sometimes called {\em implicit triggers},
as opposed to the explicit triggers of Section~\ref{lazyexample}.
Triggers implement by-need computation (i.e., lazy execution)
and are installed with the \OzInline{ByNeed} operation.
We will not say much about triggers in this article.
For more information, see~\cite{book,futures}.

The thread store consists of a set of threads.
Each thread is defined by a statement $S_i$.
Threads can only have references in the constraint store,
not into the other stores.
This means that
the only way for threads to communicate and synchronize
is through shared references in the constraint store.
We say a thread is {\em runnable}, also called {\em ready},
if it can execute its statement.
Threads are {\em dataflow} threads, i.e.,
a thread becomes runnable
when the arguments needed by its statement are bound.
If an argument is unbound
then the thread automatically suspends until
the argument is bound.
Since the constraint store is monotonic,
a thread that is runnable
will stay runnable at least until
it executes one step of its statement.
The system guarantees weak fairness, which
implies that a runnable thread will eventually execute.

\begin{figure}
\figrule
\begin{tabular}{@{}lr@{~}l@{~}l@{}}
\multicolumn{2}{@{}l@{~}}{\nt{s} ::=} & \OzInline{\OzKeyword{skip}} & \\
 & $|$ & \nt{x}${}_1$=\nt{x}${}_2$ & \\
 & $|$ & \nt{x}=\nt{l}(\nt{f}${}_1$:\nt{x}${}_1$ ...
                       \nt{f}${}_n$:\nt{x}${}_n$)
                & \\
 & $|$ & \nt{s}${}_1$ \nt{s}${}_2$ & \\
 & $|$ & \OzInline{\OzKeyword{local}} \nt{x} \OzInline{\OzKeyword{in}} \nt{s} \OzInline{\OzKeyword{end}} & \\
 & $|$ & \OzInline{\OzKeyword{if}} \nt{x} \OzInline{\OzKeyword{then}} \nt{s}${}_1$ \OzInline{\OzKeyword{else}} \nt{s}${}_2$ \OzInline{\OzKeyword{end}}
 & \\
 & $|$ & \OzInline{\OzKeyword{case}} \nt{x} \OzInline{\OzKeyword{of}} 
   \nt{l}(\nt{f}${}_1$:\nt{x}${}_1$ ... \nt{f}${}_n$:\nt{x}${}_n$)
   \OzInline{\OzKeyword{then}} \nt{s}${}_1$ \OzInline{\OzKeyword{else}} \nt{s}${}_2$ \OzInline{\OzKeyword{end}}
 & \\
 & $|$ & \OzInline{\OzKeyword{proc}\OzSpace{1}\OzChar\{}\nt{x} \nt{y}${}_1$ ... \nt{y}${}_n$\OzInline{\OzChar\}} \nt{s} \OzInline{\OzKeyword{end}} & \\
 & $|$ & \OzInline{\OzChar\{}\nt{x} \nt{y}${}_1$ ... \nt{y}${}_n$\OzInline{\OzChar\}}  & {\em CORE} \\
 &     & \\
 & $|$ & \OzInline{\OzKeyword{thread}} \nt{s} \OzInline{\OzKeyword{end}} & {\em CONCURRENCY} \\
 &     & \\
 & $|$ & \OzInline{\OzChar\{ByNeed} \nt{x} \nt{y}\OzInline{\OzChar\}} & {\em LAZINESS} \\
 &     & \\
 & $|$ & \OzInline{\OzKeyword{try}} \nt{s}${}_1$ \OzInline{\OzKeyword{catch}} \nt{x} \OzInline{\OzKeyword{then}} \nt{s}${}_2$ \OzInline{\OzKeyword{end}} & \\
 & $|$ & \OzInline{\OzKeyword{raise}} \nt{x} \OzInline{\OzKeyword{end}}  & {\em EXCEPTIONS} \\
 &     & \\
 & $|$ & \OzInline{\OzChar\{NewName} \nt{x}\OzInline{\OzChar\}} & {\em SECURITY} \\
 &     & \\
 & $|$ & \OzInline{\OzChar\{IsDet} \nt{x} \nt{y}\OzInline{\OzChar\}} & \\
 & $|$ & \OzInline{\OzChar\{NewCell} \nt{x} \nt{y}\OzInline{\OzChar\}} & \\
 & $|$ & \OzInline{\OzChar\{Exchange} \nt{x} \nt{y} \nt{z}\OzInline{\OzChar\}} & {\em STATE} \\
 &     & \\
 & $|$ & \nt{space} & {\em SEARCH} \\
\end{tabular}
\caption{The Oz kernel language.}\label{kernel}
\figrule
\end{figure}

\subsection{The kernel language}
\label{kernellang}
\label{lazykernel}

All Oz execution can be defined
in terms of a simple kernel language,
whose syntax is defined in Figure~\ref{kernel}.
The full Oz language provides syntactic
support for additional language entities
(such as functions, ports, objects, classes, and functors).
The system hides their efficient implementation
while respecting their definitions
in terms of the kernel language.
This performance optimization
can be seen as a second kernel language,
in between full Oz and the kernel language.
The second kernel language is implemented directly.

From the kernel language viewpoint,
$n$-ary functions are just $(n+1)$-ary procedures,
where the last argument is the function's output.
In Figure~\ref{kernel},
statements are denoted by \nt{s},
computation space operations by
\nt{space} (see Figure~\ref{compspacesfig}),
logic variables by \nt{x}, \nt{y}, \nt{z},
record labels by \nt{l},
and record field names by \nt{f}.

The semantics of the kernel language is given
in~\cite{book} (except for spaces)
and~\cite{Schulte:02,SchulteDiss:00} (for spaces).
For comparison,
the semantics of the original Oz language
is given in~\cite{kerneloz}.
The kernel language splits naturally into seven parts:
\begin{itemize}
\item {\em CORE}: The core is
strict functional programming over a constraint store.
This is exactly deterministic logic programming with
explicit sequential control.
The \OzInline{\OzKeyword{if}} statement expects a boolean argument
(\OzInline{\OzKeyword{true}} or \OzInline{\OzKeyword{false}}).
The \OzInline{\OzKeyword{case}} statement does pattern matching.
The \OzInline{\OzKeyword{local}} statement introduces new variables
(\OzInline{\OzKeyword{declare}} is a syntactic variant whose scope extends
over the whole program).

\item {\em CONCURRENCY}:
The concurrency support adds explicit thread creation.
Together with the core, this gives {\em dataflow concurrency},
which is a form of declarative concurrency.
Compared to a sequential program,
this gives the same results
but incrementally instead of all at once.
This is deterministic logic programming with
more flexible control than the core alone.
This is discussed at length in~\cite{book}.

\item {\em LAZINESS}:
The laziness support adds
the \OzInline{ByNeed} operation, which allows to
express lazy execution, which is the basic idea
of nonstrict functional languages
such as Haskell~\cite{futures,Mehl:diss,haskell}.\footnote{In Mozart,
the module \OzInline{Value} contains this operation: \OzInline{ByNeed=Value.byNeed}.}
Together with the core, this gives {\em demand-driven concurrency},
which is another form of declarative concurrency.
Lazy execution gives the same results as eager execution,
but calculates only what is needed to achieve the results.
Again, this is deterministic logic programming with
more flexible control than the core alone.
This is important for resource management and program modularity.
Lazy execution can give results in cases
when eager execution does not terminate.

\item {\em EXCEPTIONS}:
The exception-handling support adds
an operation, \OzInline{\OzKeyword{try}}, to create an exception context
and an operation, \OzInline{\OzKeyword{raise}}, to jump to the innermost
enclosing exception context.

\item {\em SECURITY}:
The security support adds {\em name values},
which are unforgeable constants that do not
have a printable representation.
Calling \OzInline{\OzChar\{NewName\OzSpace{1}X\OzChar\}} creates a fresh name
and binds it to \OzInline{X}.
A name is a first-class ``right'' or ``key''
that supports many programming techniques
related to security and encapsulation.

\item {\em STATE}:
The state support adds explicit cell creation
and an exchange operation, which atomically
reads a cell's content and replaces it with a new content.
This is sufficient for
sequential object-oriented programming~\cite{opm,HenzDiss:97,HenzOFCCP:97}.
Another, equivalent way to add state
is by means of ports, which are explained in Section~\ref{state}.

\item {\em SEARCH}:
The search support adds operations on computation spaces
(shown as \nt{space}), which are explained in Section~\ref{compspaces}.
This allows to express nondeterministic logic programming
(see Sections~\ref{nondetlp} and~\ref{moresearch}).
A computation space encapsulates a choice point,
i.e., don't-know nondeterminism,
allowing the program to decide how to pick alternatives.
Section~\ref{compspaces} explains spaces in
more detail and shows how to program search with them.
The \OzInline{\OzKeyword{choice}} statement, which is used in the examples
of Sections~\ref{nondetlp} and~\ref{aggreg},
can be programmed with spaces (see Section~\ref{andorradis}).


\end{itemize}

\subsubsection{Concurrency and state}

Adding both concurrency and state to the core
results in the most expressive computation model.
There are two basic approaches to program in it:
message passing with active objects or 
atomic actions on shared state.
Active objects are used in Erlang~\cite{erlang}.
Atomic actions are used in Java
and other concurrent object-oriented languages~\cite{lea2}.
These two approaches have the same expressive power,
but are appropriate for different classes of applications
(multi-agent versus data-centered)~\cite{book,needhamlauer78}.

\subsubsection{Nondeterministic choice}

Concurrent logic programming is obtained by
extending the core with concurrency and nondeterministic choice.
This gives a model that is more expressive than
declarative concurrency and less expressive
than concurrency and state used together.
Nondeterministic choice means to wait concurrently
for one of several conditions to become true.
For example, we could add the operation \OzInline{WaitTwo}
to the core with concurrency.
\OzInline{\OzChar\{WaitTwo\OzSpace{1}X\OzSpace{1}Y\OzChar\}} blocks until either \OzInline{X} or \OzInline{Y}
is bound to a nonvariable term.\footnote{In Mozart,
the module \OzInline{Record} contains this operation:
\OzInline{\OzChar\{WaitTwo\OzSpace{1}X\OzSpace{1}Y\OzChar\}} is written as \OzInline{\OzChar\{Record.waitOr\OzSpace{1}X\OzChar\#Y\OzChar\}}.}
It then returns with 1 or 2.
It can return 1 if \OzInline{X} is bound and 2 if \OzInline{Y} is bound.
\OzInline{WaitTwo} does not need to be added as an additional concept;
it can be programmed in the core with concurrency and state.

\subsubsection{Lazy functions}

The \OzInline{lazy} annotation used in Section~\ref{lazyexample}
is defined in terms of \OzInline{ByNeed}.
Calling \OzInline{\OzChar\{ByNeed\OzSpace{1}P\OzSpace{1}X\OzChar\}}
adds the trigger \OzInline{(X,P)} to the trigger store.
This makes \OzInline{X} behave as a read-only variable.
Doing a computation that needs \OzInline{X} or attempts to bind \OzInline{X}
will block the computation,
execute \OzInline{\OzChar\{P\OzSpace{1}Y\OzChar\}} in a new thread,
bind \OzInline{Y} to \OzInline{X}, and then continue.
We say a value is {\em needed} by an operation if the
thread executing the operation would suspend if the
value were not present.
For example, the function:
\begin{oz2texdisplay}\OzSpace{3}\OzKeyword{fun}\OzSpace{1}lazy\OzSpace{1}\OzChar\{Generate\OzSpace{1}N\OzChar\}\OzEol
\OzSpace{6}N|\OzChar\{Generate\OzSpace{1}N+1\OzChar\}\OzEol
\OzSpace{3}\OzKeyword{end}\end{oz2texdisplay}
is defined as:
\newpage
\begin{oz2texdisplay}\OzSpace{3}\OzKeyword{fun}\OzSpace{1}\OzChar\{Generate\OzSpace{1}N\OzChar\}\OzEol
\OzSpace{3}P\OzSpace{1}X\OzSpace{1}\OzKeyword{in}\OzEol
\OzSpace{6}\OzKeyword{proc}\OzSpace{1}\OzChar\{P\OzSpace{1}Y\OzChar\}\OzSpace{1}Y=N|\OzChar\{Generate\OzSpace{1}N+1\OzChar\}\OzSpace{1}\OzKeyword{end}\OzEol
\OzSpace{6}\OzChar\{ByNeed\OzSpace{1}P\OzSpace{1}X\OzChar\}\OzEol
\OzSpace{6}X\OzEol
\OzSpace{3}\OzKeyword{end}\end{oz2texdisplay}
\OzInline{P} will only be called when
the value of \OzInline{\OzChar\{Generate\OzSpace{1}N\OzChar\}} is needed.
We make two comments about this definition.
First, the \OzInline{lazy} annotation is given explicitly by the programmer.
Functions without it are eager.
Second, Mozart threads are extremely lightweight,
so the definition is practical.
This is a different approach than in nonstrict languages such as Haskell,
where lazy evaluation is the default and strictness analysis
is used to regain the efficiency
of eager evaluation~\cite{haskell}.

\begin{figure}
\figrule
\psfig{file=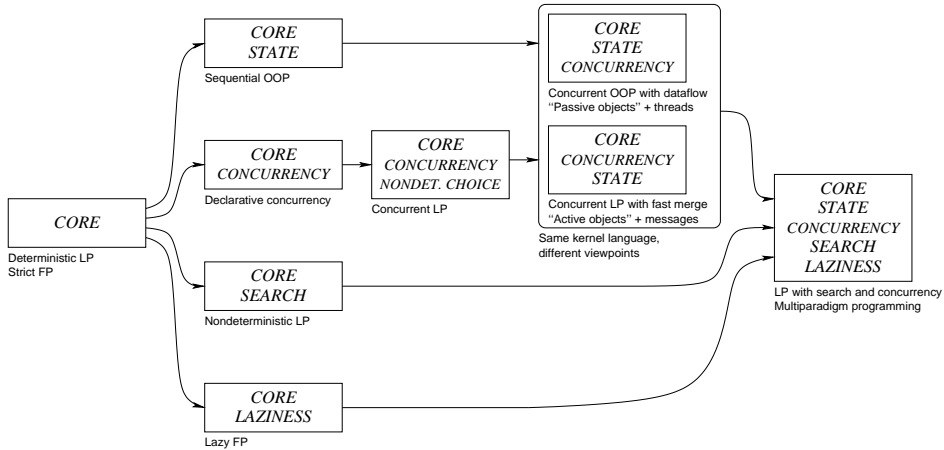,width=\textwidth}
\caption{Some programming paradigms in Oz.}\label{paradigms}
\figrule
\end{figure}

\subsection{Multiparadigm programming}
\label{multiparadigm}

Many different programming styles or ``paradigms'' are possible
by limiting oneself to different subsets of the kernel language.
Some popular styles are object-oriented programming
(programming with state, encapsulation, and inheritance),
functional programming (programming with values and pure functions),
constraint programming (programming with deduction and search),
and sequential programming
(programming with a totally-ordered sequence of instructions).
Some interesting subsets of the kernel language
are shown in Figure~\ref{paradigms}.
The full Oz language provides
syntactic and implementation support that makes
these paradigms and many others equally easy to use.
The execution model is simple and general, which allows
the different styles to coexist comfortably.
This ability is known as {\em multiparadigm} programming.

The justification of limiting oneself to one particular paradigm
is that the program may be easier to write or reason about.
For example, if the \OzInline{\OzKeyword{thread}} construct is not
used, then the program is purely sequential.
If the \OzInline{ByNeed} operation is not used, then the program is strict.
Experience shows that
different levels of abstraction often need different
paradigms (see Section~\ref{supportmp})~\cite{ftsrparadigm,book}.
Even if the same basic functionality is provided,
it may be useful to view it according to different
paradigms depending on the application needs~\cite{needhamlauer78}.

How is it possible for such a simple kernel language to
support such different programming styles?
It is because paradigms have many concepts in common,
as Figures~\ref{kernel} and~\ref{paradigms} show.
A good example is sequential object-oriented programming,
which can be built
from the core by adding just state
(see~\cite{opm} for details):
\begin{itemize}
\item Procedures behave as objects when they internally reference state.
\item Methods are different procedures that reference the {\em same} state.
\item Classes are records that group related method definitions.
\item Inheritance is an operation that takes a set
of method definitions and one or more class records,
and constructs a new class record.
\item Creation of new object instances is done by
a higher-order procedure that
takes a class record and associates
a new state pointer with it.
\end{itemize}
Oz has syntactic support to make this style easy to use
and implementation support to make it efficient.
The same applies to the declarative paradigms
of functional and logic programming.
Strict functions are restricted versions of procedures
in which the binding is directional.
Lazy functions are implemented with \OzInline{ByNeed}.

For logic programming,
procedures become relations when they have
a logical semantics in addition to their
operational semantics.
This is true within the core.
It remains true if one adds concurrency and laziness to the core.
We illustrate the logical semantics
with many examples in this article, starting in Section~\ref{detlp}.
In the core,
the \texttt{if} and \texttt{case} statements have
a logical semantics, i.e., they check entailment and disentailment.
To make the execution {\em complete}, i.e.,
to always find a constructive proof when one exists,
it is necessary to add search.
Oz supports search by means of computation spaces.
When combined with the rest of the model,
they make it possible to program a wide variety
of search algorithms in Oz,
as explained in the next section.

\subsection{Computation spaces}
\label{compspaces} 

Computation spaces are a powerful abstraction
that permits the high-level programming
of search abstractions and deep guard combinators,
both of which are important for
constraint and logic programming.
Spaces are a natural way to integrate search
into a concurrent system.
Spaces can be implemented efficiently:
on real-world problems
the Mozart 1.1.0 implementation
using copying and recomputation
is competitive in time and memory use
with traditional systems
using trailing-based backtracking~\cite{Schulte:ICLP99}.
Spaces are compositional,
i.e., they can be nested,
which is important for building
well-structured programs.


This section defines computation spaces,
the operations that can be performed on them
(see Figure~\ref{compspacesfig}),
and gives a few examples
of how to use them to program search.
The discussion in this section
follows the model in~\cite{Schulte:02,SchulteDiss:00}.
This model is implemented in Mozart 1.1.0~\cite{mozart110}
and refines the one presented
in the articles~\cite{Engines:97,Schulte:00}.
The space abstraction can be made language-independent;
\cite{figaro-parimplws99} describes a C++ implementation
of a similar abstraction
that supports both trailing and copying.

\begin{figure}
\figrule
\psfig{file=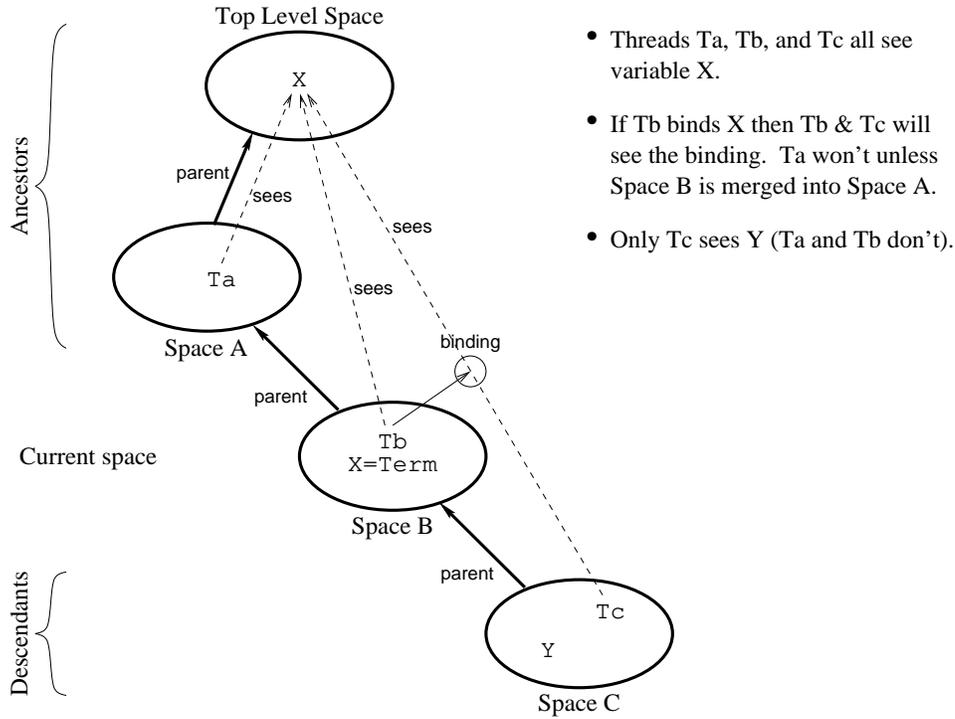,width=\textwidth}
\caption{Visibility of variables and bindings in nested spaces.}
\label{spacehierarchy}
\figrule
\end{figure}

\subsubsection{Definition}

A computation space is just an Oz store with its four parts.
The store we have seen so far is a single computation space
with equality constraints over rational trees.
To deal with search, we extend this in two ways.
First, we allow spaces to be nested.
Second, we allow other constraint systems in a space.
Since spaces are used to encapsulate
potential variable bindings, it is important
to be precise about the visibility of variables and bindings.
Figure~\ref{spacehierarchy} gives an example.
The general rules
for the structure of computation spaces are as follows:
\begin{itemize}
\item{}There is always a {\em top level} computation space
where threads may interact with the external world.
The top level space is just the store of Section~\ref{store}.
Because the top level space interacts with the external world,
its constraint store always remains consistent, that is,
each variable has at most one binding that never changes
once it is made.
A thread that tries
to add an inconsistent binding
to the top level constraint store
will raise a failure exception.

\item{}A thread may create a new computation space.
The new space is called a {\em child space}.
The current space is the child's {\em parent space}.
At any time, there is a tree of computation spaces
in which the top level space is the root.
With respect to a given space,
a higher one in the tree (closer to the root)
is called an {\em ancestor} and
a lower one is called a {\em descendant}.

\item{}A thread always belongs to exactly one computation space.
A variable always belongs to exactly one computation space.

\item{}A thread sees and may access variables
belonging to its space as well as to all ancestor spaces.
The thread cannot see the variables of descendant spaces.

\item{}A thread cannot see the variables of a child space,
unless the child space is {\em merged} with its parent.
Space merging is an explicit program operation.
It causes the child space to disappear and all the child's content
to be added to the parent space.

\item{}A thread may add bindings to variables visible to it.
This means that it may bind variables belonging to its
space or to its ancestor spaces.
The binding will only be visible in
the current space and its descendants.
That is, the parent space does not see the binding unless
the current space is merged with it.

\item{}If a thread in a child space tries
to add an inconsistent binding to its constraint store,
then the space fails.

\end{itemize}

\subsubsection{State of a space}

A space is {\em runnable} if it or a descendant contains
a runnable thread, and {\em blocked} otherwise.
Let us run all threads in the space and its
descendants, until the space is blocked.
Then the space can be in one of
the following further states:
\begin{itemize}
\item The space is {\em stable}.
This means that no additional
bindings done in an ancestor can make the space runnable.
A stable space can be in four further states:
\begin{itemize}
\item The space is {\em succeeded}.
This means that it contains no choice points.
A succeeded space contains a solution.

\item The space is {\em distributable}.
This means that the space has one thread
that is suspended on a choice point
with two or more alternatives.
A space can have at most one choice point;
attempting to create another gives a run-time error.

\item The space is {\em failed}.  This is defined
in the previous section; it means that the space
attempted to bind the same variable to two different values.
No further execution happens in the space.

\item The space is {\em merged}.
This means that the space has been discarded
and its constraint store has been added to its parent.
Any further operation on the space is an error.
This state is the end of a space's lifetime.

\end{itemize}

\item The space is {\em suspended}.
This means that additional bindings
done in an ancestor can make the space runnable.
Being suspended is usually a temporary condition
due to concurrency.
It means that some ancestor space has not yet
transferred all required information to the space.
A space that stays suspended indefinitely
usually indicates a programmer error.
\end{itemize}

\subsubsection{Programming search}
\label{spaceops}

A {\em search strategy} defines how the search tree is explored,
e.g., depth-first seach, limited discrepancy search,
best-first search, and branch-and-bound search.
A {\em distribution strategy} defines
the shape and content of the search tree,
i.e., how many alternatives exist at a node and
what constraint is added for each alternative.
Computation spaces can be used to program
search strategies and distribution strategies
independent of each other.
That is, any search strategy can be used together
with any distribution strategy.
Here is how it is done:
\begin{itemize}
\item Create the space and initialize it
by running an internal program that defines all
the variables and constraints in the space.

\item Propagate information inside the space.
The constraints in a space have an operational semantics.
In Oz terminology, an operationalized version of
a constraint is called a {\em propagator}.
Propagators execute concurrently; each propagator
executes inside its own thread.
Each propagator reads its arguments and attempts to add
information to the constraint store by restricting
the domains of its arguments.

\item All propagators execute until no more information
can be added to the store in this manner.
This is a fixpoint calculation.
When no more information can be added,
then the fixpoint is reached and
the space has become stable.

\item During a space's execution,
the computation inside the space
can decide to create a choice point.
The decision which constraint to add
for each alternative defines the distribution strategy.
One of the space's threads will suspend when
the choice point is created.

\item When the space has become stable,
then execution continues outside the space,
to decide what to do next.
There are different possibilities depending on whether or not
a choice point has been created in the space.
If there is none, then execution can stop and return with a solution.
If there is one, 
then the search strategy decides which alternative
to choose and commits to that alternative.
\end{itemize}
Notice that the distribution strategy is problem-dependent:
to add a constraint we need to know the problem's constraints.
On the other hand,
the search strategy is problem-independent:
to pick an alternative we do not need to know
which constraint it corresponds to.
The next section explains
the operations we need to implement this approach.
Then, Section~\ref{spaceexamples}
gives some examples of how to program search.

\begin{figure}
\figrule
\begin{center}
\begin{tabular}{lrl}
\multicolumn{2}{l}{\nt{space} ::=}
       & \OzInline{\OzChar\{NewSpace} \nt{x} \nt{y}\OzInline{\OzChar\}} \\
 & $|$ & \OzInline{\OzChar\{Choose} \nt{x} \nt{y}\OzInline{\OzChar\}} \\
 & $|$ & \OzInline{\OzChar\{Ask} \nt{x} \nt{y}\OzInline{\OzChar\}} \\
 & $|$ & \OzInline{\OzChar\{Commit} \nt{x} \nt{y}\OzInline{\OzChar\}} \\
 & $|$ & \OzInline{\OzChar\{Clone} \nt{x} \nt{y}\OzInline{\OzChar\}} \\
 & $|$ & \OzInline{\OzChar\{Inject} \nt{x} \nt{y}\OzInline{\OzChar\}} \\
 & $|$ & \OzInline{\OzChar\{Merge} \nt{x} \nt{y}\OzInline{\OzChar\}}
\end{tabular}
\end{center}
\caption{Primitive operations for computation spaces.}
\label{compspacesfig}
\figrule
\end{figure}

\subsubsection{Space operations}

Now we know enough to define the primitive space operations.
There are seven principal ones (see Figure~\ref{compspacesfig}).
\begin{itemize}
\item \OzInline{\OzChar\{NewSpace\OzSpace{1}P\OzSpace{1}X\OzChar\}}, when given a unary
procedure \OzInline{P}, creates a new computation space \OzInline{X}.
In this space, a fresh variable \OzInline{R}, called the
{\em root variable}, is created,
and \OzInline{\OzChar\{P\OzSpace{1}R\OzChar\}} is invoked in a new thread.

\item \OzInline{\OzChar\{Choose\OzSpace{1}N\OzSpace{1}Y\OzChar\}} is the only operation
that executes {\em inside} a space.
It creates a choice point with \OzInline{N} alternatives.
Then it blocks, waiting for an alternative to be chosen
by a \OzInline{Commit} operation on the space (see below).
The \OzInline{Choose} call defines only the {\em number} of 
alternatives; it does not
specify what to do for any given alternative.
\OzInline{Choose} returns with \OzInline{Y=I} when
alternative \OzInline{1}$\leq$\OzInline{I}$\leq$\OzInline{N} is chosen.
A maximum of one choice point
may exist in a space at any time.

\item \OzInline{\OzChar\{Ask\OzSpace{1}X\OzSpace{1}A\OzChar\}} asks the space \OzInline{X} for its status.
As soon as the space becomes stable, \OzInline{A} is bound.
If \OzInline{X} is failed, merged, or succeeded,
then \OzInline{A} is bound to \OzInline{failed}, \OzInline{merged}, or \OzInline{succeeded}.
If \OzInline{X} is distributable, then \OzInline{A=alternatives(N)},
where \OzInline{N} is the number of alternatives.

\item \OzInline{\OzChar\{Commit\OzSpace{1}X\OzSpace{1}I\OzChar\}}, if \OzInline{X} is a distributable space,
causes the blocked \OzInline{Choose} call in the space to continue
with \OzInline{I} as its result.
This may cause a stable space to become not stable again.
The space will resume execution until a new fixpoint is reached.
The integer \OzInline{I} must satisfy \OzInline{1}$\leq$\OzInline{I}$\leq$\OzInline{N},
where \OzInline{N} is the first argument of the \OzInline{Choose} call.

\item \OzInline{\OzChar\{Clone\OzSpace{1}X\OzSpace{1}C\OzChar\}}, if \OzInline{X} is a stable space,
creates an identical copy (a {\em clone}) of \OzInline{X} in \OzInline{C}.
This allows the alternatives of a distributable space to
be explored independently.

\item \OzInline{\OzChar\{Inject\OzSpace{1}X\OzSpace{1}P\OzChar\}} is similar to space creation
except that it uses an existing space \OzInline{X}.
It creates a new thread in the space
and invokes \OzInline{\OzChar\{P\OzSpace{1}R\OzChar\}} in the thread,
where \OzInline{R} is the space's root variable.
This may cause a stable space to become not stable again.
The space will resume execution until a new fixpoint is reached.
Adding constraints to an existing space is necessary
for some search strategies such as branch-and-bound
and saturation.

\item \OzInline{\OzChar\{Merge\OzSpace{1}X\OzSpace{1}Y\OzChar\}} binds \OzInline{Y} to the 
root variable of space \OzInline{X} and discards the space.

\end{itemize}

\begin{figure}
\figrule
\psfig{file=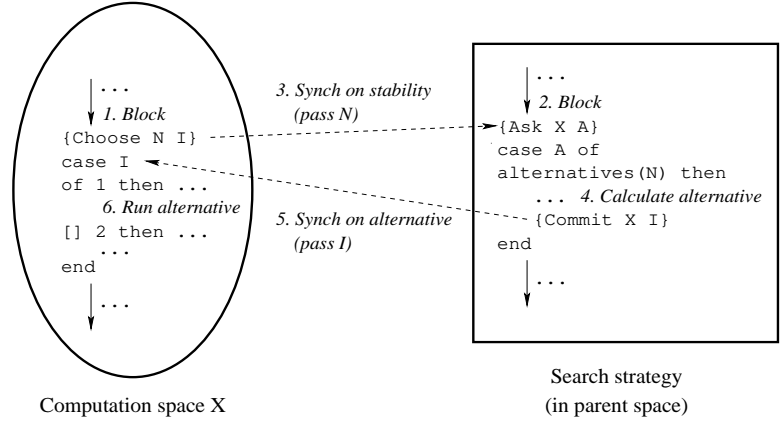,width=\textwidth}
\caption{Communication between a space and its search strategy.}
\label{spacesynch}
\figrule
\end{figure}

\subsubsection{Using spaces}
\label{spaceexamples}

These seven primitive operations are enough to
define many search strategies and distribution strategies.
The basic technique is to use \OzInline{Choose}, \OzInline{Ask},
and \OzInline{Commit} to communicate between
the inside of the space and the outside of the space.
Figure~\ref{spacesynch} shows how the communication
works: first the space informs the search strategy
of the total number of alternatives (\OzInline{N}).
Then the search strategy picks one (\OzInline{I}) and
informs the space.
Let us now present briefly a few examples of how to use spaces.
For complete information on these examples
and many other examples we refer the reader
to~\cite{Schulte:02,SchulteDiss:00}.

\paragraph{Depth-first search.}

Our first example implements a search strategy.
Figure~\ref{DFS} shows how to program
depth-first single solution search
in the case of binary choice points.
This explores the search tree in depth-first
manner and returns the first solution it finds.
The problem is defined as a unary procedure \OzInline{\OzChar\{P\OzSpace{1}Sol\OzChar\}}
that gives a reference to the solution \OzInline{Sol},
just like the example in Section~\ref{scalable}.
The solution is returned
in a one-element list as \OzInline{[Sol]}.
If there is no solution, then \OzInline{nil} is returned.
In \OzInline{P}, choice points are defined with the \OzInline{Choose} operation.

\begin{figure}
\figrule
\begin{center}
\begin{oz2texdisplay}\OzSpace{3}\OzKeyword{fun}\OzSpace{1}\OzChar\{DFE\OzSpace{1}S\OzChar\}\OzEol
\OzSpace{6}\OzKeyword{case}\OzSpace{1}\OzChar\{Ask\OzSpace{1}S\OzChar\}\OzEol
\OzSpace{6}\OzKeyword{of}\OzSpace{1}failed\OzSpace{1}\OzKeyword{then}\OzSpace{1}nil\OzEol
\OzSpace{6}[]\OzSpace{1}succeeded\OzSpace{1}\OzKeyword{then}\OzSpace{1}[S]\OzEol
\OzSpace{6}[]\OzSpace{1}alternatives(2)\OzSpace{1}\OzKeyword{then}\OzSpace{1}C=\OzChar\{Clone\OzSpace{1}S\OzChar\}\OzSpace{1}\OzKeyword{in}\OzEol
\OzSpace{9}\OzChar\{Commit\OzSpace{1}S\OzSpace{1}1\OzChar\}\OzEol
\OzSpace{9}\OzKeyword{case}\OzSpace{1}\OzChar\{DFE\OzSpace{1}S\OzChar\}\OzSpace{1}\OzKeyword{of}\OzSpace{1}nil\OzSpace{1}\OzKeyword{then}\OzSpace{1}\OzChar\{Commit\OzSpace{1}C\OzSpace{1}2\OzChar\}\OzSpace{1}\OzChar\{DFE\OzSpace{1}C\OzChar\}\OzEol
\OzSpace{9}[]\OzSpace{1}[T]\OzSpace{1}\OzKeyword{then}\OzSpace{1}[T]\OzEol
\OzSpace{9}\OzKeyword{end}\OzEol
\OzSpace{6}\OzKeyword{end}\OzEol
\OzSpace{3}\OzKeyword{end}\OzEol
\OzEol
\OzSpace{3}\OzEolComment{\OzSpace{1}Given\OzSpace{1}procedure\OzSpace{1}\OzChar\{P\OzSpace{1}Sol\OzChar\},\OzSpace{1}returns\OzSpace{1}solution\OzSpace{1}[Sol]\OzSpace{1}or\OzSpace{1}nil:}\OzSpace{3}\OzKeyword{fun}\OzSpace{1}\OzChar\{DFS\OzSpace{1}P\OzChar\}\OzEol
\OzSpace{6}\OzKeyword{case}\OzSpace{1}\OzChar\{DFE\OzSpace{1}\OzChar\{NewSpace\OzSpace{1}P\OzChar\}\OzChar\}\OzSpace{1}\OzKeyword{of}\OzSpace{1}nil\OzSpace{1}\OzKeyword{then}\OzSpace{1}nil\OzEol
\OzSpace{6}[]\OzSpace{1}[S]\OzSpace{1}\OzKeyword{then}\OzSpace{1}[\OzChar\{Merge\OzSpace{1}S\OzChar\}]\OzEol
\OzSpace{6}\OzKeyword{end}\OzEol
\OzSpace{3}\OzKeyword{end}\end{oz2texdisplay}
\end{center}
\caption{Depth-first single solution search.}
\label{DFS}
\figrule
\end{figure}

\paragraph{Naive choice point.}

Our second example implements a distribution strategy.
Let us implement a naive choice point,
namely one that defines a set of alternative statements
to be chosen.
This can be defined as follows:
\newpage
\begin{oz2texdisplay}\OzSpace{3}\OzKeyword{case}\OzSpace{1}\OzChar\{Choose\OzSpace{1}N\OzChar\}\OzEol
\OzSpace{3}\OzKeyword{of}\OzSpace{1}1\OzSpace{1}\OzKeyword{then}\OzSpace{1}$S_1$\OzEol
\OzSpace{3}[]\OzSpace{1}2\OzSpace{1}\OzKeyword{then}\OzSpace{1}$S_2$\OzEol
\OzSpace{3}...\OzEol
\OzSpace{3}[]\OzSpace{1}N\OzSpace{1}\OzKeyword{then}\OzSpace{1}$S_n$\OzEol
\OzSpace{3}\OzKeyword{end}\end{oz2texdisplay}
Oz provides the following more convenient syntax
for this technique:
\begin{oz2texdisplay}\OzSpace{3}\OzKeyword{choice}\OzSpace{1}$S_1$\OzSpace{1}[]\OzSpace{1}...\OzSpace{1}[]\OzSpace{1}$S_n$\OzSpace{1}\OzKeyword{end}\end{oz2texdisplay}
This is exactly how the \OzInline{\OzKeyword{choice}} statement is defined.
This statement can be used with any search strategy,
such as the depth-first strategy we defined previously
or other strategies.

\paragraph{Andorra-style disjunction (the \OzInline{\OzKeyword{dis}} statement).}
\label{andorradis}

Let us now define a slightly more complex distribution strategy.
We define the \OzInline{\OzKeyword{dis}} statement, which is an extension of \OzInline{\OzKeyword{choice}} that
eliminates failed alternatives and commits
immediately if there is a single remaining alternative:
\begin{oz2texdisplay}\OzSpace{3}\OzKeyword{dis}\OzSpace{1}$G_1$\OzSpace{1}\OzKeyword{then}\OzSpace{1}$S_1$\OzSpace{1}[]\OzSpace{1}...\OzSpace{1}[]\OzSpace{1}$G_n$\OzSpace{1}\OzKeyword{then}\OzSpace{1}$S_n$\OzSpace{1}\OzKeyword{end}\end{oz2texdisplay}
In contrast to \OzInline{\OzKeyword{choice}}, each alternative of a \OzInline{\OzKeyword{dis}} statement
has both a guard and a body.
The guards are used immediately to check failure.
If a guard $G_i$ fails then its alternative is eliminated.
This extension is sometimes called determinacy-directed execution.
It was discovered by D.H.D. Warren
and called the {\em Andorra principle}~\cite{andorra,andorramodel}.

The \OzInline{\OzKeyword{dis}} statement can be
programmed with the space operations as follows.
First encapsulate each guard
of the \OzInline{\OzKeyword{dis}} statement in a separate space.
Then execute each guard until it is stable.
Discard all failed guards.
Finally, using the \OzInline{Choose} operation,
create a choice point for the remaining guards.
See~\cite{Schulte:02,SchulteDiss:00} for details of the implementation.
It can be optimized to do first-argument indexing
in a similar way to Prolog systems.
We emphasize that
the whole implementation is written within the language.

\paragraph{The first-fail strategy.}

In practice, \OzInline{\OzKeyword{dis}} is not strong enough for solving
real constraint problems.
It is too static: its alternatives are
defined textually in the program code.
A more sophisticated distribution strategy would
look more closely at the actual state of the execution.
For example, the {\em first-fail} strategy
for finite domain constraints
looks at all variables and places a choice
point on the variable whose domain is the smallest.
First-fail can be implemented with \OzInline{Choose}
and a set of reflective operations on finite domain constraints.
The Mozart system provides first-fail
as one of many preprogrammed strategies.

\paragraph{Deep guard combinators.}

A {\em constraint combinator} is an operator that takes
constraints as arguments and combines them to form another constraint.
Spaces are a powerful way to implement constraint combinators.
Since spaces are compositional, the resulting constraints
can themselves be used as inputs to other constraint combinators.
For this reason,
these combinators are called {\em deep guard} combinators.
This is more powerful than other techniques, such as reification,
which are {\em flat}: their input constraints are limited to
simple combinations of built-in constraints.
Some examples of deep guard combinators 
that we can program are deep negation,
generalized reification,
propagation-based disjunction (such as \OzInline{\OzKeyword{dis}}),
constructive disjunction,
and deep committed-choice.

\section{Related work}
\label{related}

We first give a brief overview of research
in the area of multiparadigm programming.
We then give a short history of Oz.

\subsection{Multiparadigm languages}



Integration of paradigms is an active area of research
that has produced a variety of different languages.
We give a brief glimpse into this area.
We do not pretend to be exhaustive; that would
be the subject of another paper.
As far as we know, there is no other language
that covers as many paradigms as Oz
in an equitable way, i.e.,
with a simple formal semantics~\cite{kerneloz,book}
and an efficient implementation~\cite{amoz,Mehl:diss,ScheidhauerDiss,Schulte:02,SchulteDiss:00}.
An early discussion of multiparadigm programming in Oz
is given in~\cite{elephant}.
It gives examples in functional, logic, and object-oriented styles.

A short-term solution to integrate different paradigms is to use
a {\em coordination model}~\cite{CarrieroGelernter:89,linda}.
The prototypical coordination model is Linda,
which provides a uniform global tuple space that
can be accessed with a small set of
basic operations (concurrent reads and writes)
from any process that is connected to it.
A Linda layer can act as ``glue'' between
languages of different paradigms.
Let us now look at more substantive solutions.


Within the imperative paradigm,
there have been several efforts to add the abilities
of functional programming.
Smalltalk has ``blocks'',
which are lexically-scoped closures~\cite{smalltalk80}.
Java has inner classes, which (with minor limitations)
are lexically-scoped closures.
Java supports the {\tt final} annotation, which
allows programming with stateless objects.
Using inner classes and {\tt final} allows
to do functional programming in Java.
However, this technique is verbose and its use
is discouraged~\cite{javabook}.
More ambitious efforts are
C++ libraries such as FC++~\cite{fc}
and language extensions
such as Pizza~\cite{pizza} and Brew~\cite{brew},
which translate into Java.
These provide much better support for functional programming.

Within the functional paradigm,
the easiest way to allow imperative programming
is to add locations with destructive assignment. 
This route was taken by languages such as
Lisp \cite{commonlisp}, Scheme \cite{R4RS:91},
and SML \cite{HarperMacQueenMilner:86}. 
The M-structures of Id \cite{Nikhil:Id:91} and its 
successor pH \cite{Nikhil:pH:94,ph} fall in this category as well.
Objective Caml is a popular object-oriented dialect of ML
that takes this approach~\cite{ocaml,ocaml2}.
Oz also takes this approach, building an object system
from a functional core by adding
the cell as its location primitive.

In Haskell, state is integrated using the monadic style of 
programming \cite{Wadler:92c,PeytonJonesWadler:93} which 
generalizes the continuation-passing style. 
Because Haskell is a nonstrict language,
it cannot easily add locations with destructive assignment.
The monadic style allows to control the sequentialization 
necessary for various kinds of side effecting
(I/O, error handling, nondeterministic choice). 
However, because it imposes a global state threading,
it has difficulties when integrated with concurrency.
See~\cite{book} for a discussion of the relative
merits of the state threading approach versus
the location approach.

Within the logic paradigm,
there have been many attempts
to add an object system~\cite{Davison:93}. 
Prominent examples are Prolog++~\cite{moss}
and SICStus Objects~\cite{sicstus}.
These approaches use locations as primitives,
much like the functional approach.

Functions have been added in several ways to logic languages.
A first approach is LIFE, which
provides functions as a kind of relation
that is called by {\em entailment},
i.e, the function call waits
until its arguments have enough information.
This delaying mechanism is called {\em residuation}~\cite{life89,jlplife,Ait-Kaci+Lincoln/88/LIFE,wildlife-handbook}.
A second approach extends the basic resolution step
to include the deterministic evaluation of functions.
This execution strategy, called {\em narrowing},
underlies the Curry language~\cite{Hanus:JLP:94,curry}.
A third approach is taken by
Lambda Prolog~\cite{NadathurMiller:95}\nocite{GabbayHoggerRobinson:95}.
It uses a more powerful logic than Horn logic as a basis for programming.
In particular, functional programming is supported 
by providing $\lambda$ terms as data structures, which are handled
by a form of higher-order unification.
A fourth approach is taken by HiLog~\cite{hilog},
which introduces a higher-order syntax that can be encoded
into the first-order predicate calculus.

The Oz approach is to provide first-class procedure values
and to consider them as constants for the purposes of unification.
This approach cleanly separates the logical aspects
from the higher-order programming aspects.
All the other approaches mentioned
are more closely tied to the resolution operation.
In addition, the Oz approach provides the full power
of lexically-scoped closures as values in the language.
Finally, Oz provides entailment checking as a separate operation,
which allows it to implement call by entailment.

Erlang is a notable example of a multiparadigm language.
It has a layered design~\cite{erlang,derlang}.
Erlang programs consist of active objects that send messages to each other.
A strict functional language
is used to program the internals of the active objects.
Each active object contains one thread that runs a recursive function.
The object state is contained in the function arguments.
This model is extended further with distribution
and fault tolerance.

The layered approach is also taken by pH,
a language designed for defining algorithms
with implicit parallelism~\cite{Nikhil:pH:94,ph}.
Its core is based on Haskell.
It has two extensions.
The first extension is a single-assignment data type, I-structures.
This allows to write functional programs that have dataflow behavior.
The second extension is a mutable data type, M-structures.
This allows stateful programs.
This design has similarities to Oz,
with logic variables being the single-assignment extension
and cells the mutable extension.

Concurrent logic programming has investigated in depth the use
of logic variables for synchronization and communication.
They are one of the most expressive mechanisms for
practical concurrent programming~\cite{BalSteinerTanenbaum:89,book}.
Since logic variables are constrained monotonically,
they can express monotonic synchronization.
This allows declarative concurrency,
which is concurrent programming with no observable nondeterminism.
The concurrent logic language Strand evolved into
the coordination language
PCN \cite{Foster:PCN-Strand:93} for imperative languages.
In the functional programming community, 
the futures of Multilisp~\cite{halstead}
and the I-structures of Id~\cite{Nikhil:Id:91} 
allow to synchronize on the result of a concurrent computation.
Both realize a restricted form of logic variable.
Finally, the Goffin project~\cite{ChakravartyEtAl:95}
uses a first-order concurrent constraint language as
a coordination language for Haskell processes.


The multiparadigm language Leda
was developed for educational purposes~\cite{leda}.
It is sequential, supports
functional and object-oriented programming,
and has basic support for backtracking and a simple
form of logic programming that is a subset of Prolog.

\begin{figure}
\figrule
\psfig{file=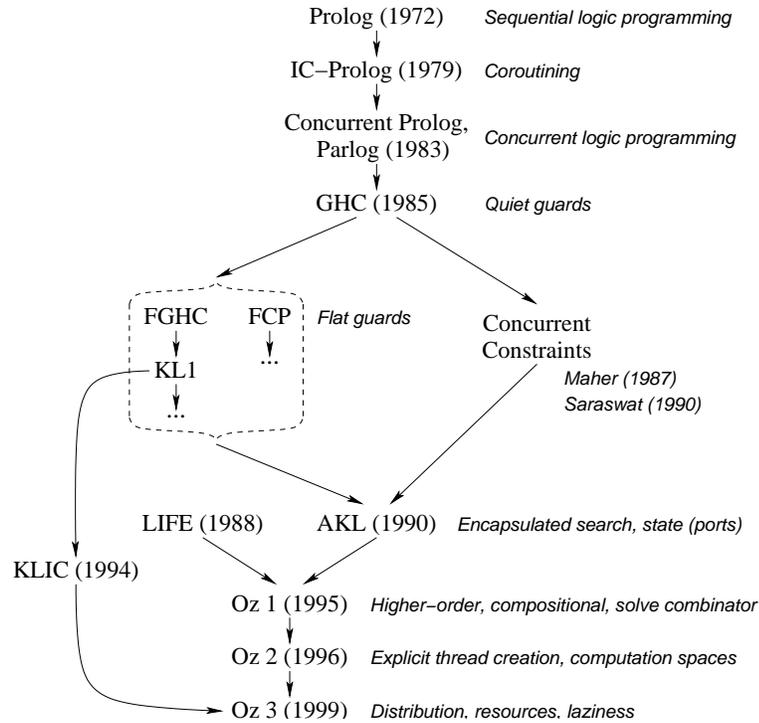,width=\textwidth}
\caption{History of Oz.}
\label{history}
\figrule
\end{figure}

\subsection{History of Oz}
\label{historysketch}

Oz is a recent descendant of a long line of logic-based languages
that originated with Prolog
(see Figure~\ref{history}).
We summarize briefly the evolutionary path
and give some of the important milestones along the way.
First experiments with concurrency were done
in the venerable IC-Prolog language
where coroutining was used
to simulate concurrent processes~\cite{CM79:,Clar82a}.
This led to Parlog and Concurrent Prolog, which introduced the process model
of logic programming, usually known as
{\em concurrent} logic programming~\cite{lncs259*30,tr003,concprolog}.
The advent of GHC (Guarded Horn Clauses) simplified
concurrent logic programming considerably by introducing the notion of
{\em quiet guards}~\cite{lncs221*168}.
A clause matching a goal will fire only if the guard is
entailed by the constraint store.
This formulation and its theoretical
underpinning were pioneered by the work of
Maher and Saraswat as they gave a solid foundation
to concurrent logic programming~\cite{maher87,saraswat90,SaraswatBook}.
The main insight is that logical notions such as
equality and entailment can be given an operational reading.
Saraswat's concurrent constraint model
is a model of concurrent programming with a logical foundation.
This model was subsequently used as the basis
for several languages including AKL and Oz.

On the practical side, systems with ``flat'' guards
(which are limited to basic constraints or system-provided tests)
were the focus of much work~\cite{deevolution}.
The flat versions of Concurrent Prolog and GHC,
called FCP and FGHC respectively,
were developed into large systems~\cite{fgcs92,Shapiro:89}.
The KL1 (Kernel Language 1) language, derived from FGHC, was implemented
in the high-performance KLIC system.
This system runs on
sequential, parallel, and distributed machines~\cite{klic}.
Some of the implementation techniques in the current Mozart system
were inspired by KLIC,
notably the distributed garbage collection algorithm.

An important subsequent development
was AKL (Andorra Kernel Language)~\cite{janson91,sverkerthesis,portpaper},
which added state (in the form of ports),
encapsulated search,
and an efficient implementation of deep guards.
AKL is the first language that combines the
abilities of constraint logic programming
and concurrent logic programming.
AKL implements encapsulated search
using a precursor of computation spaces.
When local propagation within a space
cannot choose between different disjuncts,
then the program can try each disjunct by cloning the computation space.

The initial Oz language, Oz 1,
was inspired by AKL and LIFE, and added higher-order procedures,
programmable search based on the solve combinator
(a less expressive precursor of
spaces~\cite{SchulteSmolkaEa:94,SchulteSmolka:94}),
compositional syntax,
and the cell primitive for mutable state~\cite{opm}.
Oz 1 features a new record data type
that was inspired by LIFE~\cite{SmolkaTreinen:94b,VMS:PLILP96}.
Concurrency in Oz 1 is implicit and based on lazy thread creation.
When a statement blocks, a new thread
is created that contains only the blocked statement.
The main thread is not suspended but
continues with the next statement.
Oz 1 features a concurrent
object system designed for lazy thread creation,
based on state threading and monitors.

Oz 2 improves on its predecessor Oz 1
with an improved concurrency model
and an improved model for encapsulated search.
Oz 2 replaces the solve combinator of Oz 1
by computation spaces.
In contrast to the solve combinator,
spaces allow programming important search strategies
such as parallel search, the Oz Explorer,
and strategies based on recomputation.
Oz 2 abandons implicit concurrency
in favor of an explicit thread creation construct.
Thread suspension and resumption are still
based on dataflow using logic variables.
Our experience shows
that explicit concurrency makes it easier
for the user to control application resources.
It allows the language to have an efficient and expressive
object system without
sequential state threading in method definitions.
It allows a simple debugging model and
it makes it easy to add exception handling to the language.

The current Oz language, Oz 3, conservatively extends Oz 2 with
support for first-class module specifications,
called {\em functors}~\cite{modules-98},
and for open, robust, distributed
programming~\cite{ngc98,dstutorial,toplas99,sendai99,perdio}.
A functor specifies a module in terms of the other modules it needs.
Distribution is transparent, i.e.,
the language semantics is unchanged
independent of how the program is distributed.
With respect to logic programming, the distributed
extension has two properties:
\begin{itemize}
\item The top level space
is efficiently distributed over multiple processes.
In particular, the top level store
is implemented by a practical algorithm for
distributed rational tree unification~\cite{toplas99}.
\item A child computation space is a stationary entity
that exists completely in one process.
Due to the communication overheads involved,
we have not found it worthwhile to distribute
one child space over multiple processes.
Constraint propagation {\em within} a child space
is therefore completely centralized.
Parallel search engines (see example in Section~\ref{scalable})
are implemented by putting
child spaces in different processes.
\end{itemize}
In all versions of Oz, concurrency is intended primarily
to model logical concurrency in the application
rather than to achieve parallelism (speedup) in the implementation.
However, the distributed implementation is
useful for parallel execution.
It is optimized to be
particularly efficient on shared-memory multiprocessors.
For that case,
we have experimented with an implementation
of interprocess communication
using shared pages between address spaces~\cite{ngc98}.

\section{Lessons learned}
\label{lessons}

One of the goals of the Oz project was to use logic programming
for real-world problems.
During the course of the project, we have tried out many
implementations and programming techniques, and built many applications.
From this experience, we have learned many lessons
both for practical logic programming and
for multiparadigm programming.
Here is a summary of the most important of these lessons.
We agree with the conclusions of Hughes, namely that
higher-order procedures are essential and
that laziness (demand-driven execution) is useful~\cite{fpmatters}.

\subsection{Be explicit (``magic'' does not work)}

\begin{itemize}
\item Provide explicit concurrency
(older concurrent logic programming systems have implicit concurrency).
This is
important for interaction with the environment, efficiency, facilitating
reasoning (e.g., for termination), and debugging.
It is also important for distributed programming.

\item Provide explicit search (Prolog has implicit search).
The majority of
Prolog programs solve algorithmic problems,
which do not need search, yet one cannot
use Prolog without learning about search.
Furthermore, for search problems
the search must be {\em very} controllable,
otherwise it does not scale to real applications.
Prolog's implicit search is much too weak;
this means that inefficient approaches such as
meta-interpreters are needed.
We conclude that Prolog's search is ineffective for
both algorithmic and search problems.

\item Provide explicit state (in C++ and Java, state is implicit,
e.g., Java variables are stateful unless declared {\tt final}).
By explicit state we mean that the language should declare
mutable references only where they are needed.
Explicit state should be used sparingly,
since it complicates reasoning about programs
and is costly to implement in a distributed system.
On the other hand,
explicit state is crucial for modularity, i.e., the ability
to change a program component without having
to change other components.

\item Provide explicit laziness
(in Haskell, laziness is implicit for all functions).
Explicitly declaring functions as lazy makes them easy to implement
and documents the programmer's intention.
This allows the system to pay for laziness only where it is used.
A second reason is declarative concurrency:
supporting it well requires eager as well as lazy functions.
A third reason is explicit state.
With implicit laziness (and a fortiori with nonstrictness),
it is harder to reason about
functions that use explicit state.
This is because the order of function evaluation
is not determined by syntax but is data dependent.

\end{itemize}

\subsection{Provide primitives for building abstractions}

\begin{itemize}
\item Full compositionality is essential:
everything can be nested everywhere.
For maximum usefulness,
this requires higher-order procedures with lexical scoping.
User-defined abstractions should be carefully designed
to be fully compositional.

\item The language should be complete enough so
that it is easy to define new abstractions.
The developer should have all the primitives necessary
to build powerful abstractions.
For example, in addition to lexical scoping, it is
important to have {\em read-only} logic variables,
which allow to build abstractions that export
logic variables and still protect them~\cite{futures}.
There is no distinction between built-in abstractions and
application-specific ones, except possibly regarding performance.
Examples of built-in abstractions are the object system,
reentrant locks, distribution support, and user interface support.
\end{itemize}

\subsection{Factorize and be lean}

Complexity is a source of problems and
must be reduced as much as possible:
\begin{itemize}
\item Factorize the design at {\em all} levels of abstraction,
both in the language and the implementation.
Keep the number of primitive operations to a minimum.
This goal is often in conflict with the goal of
having an efficient implementation.
Satisfying both is difficult, but sometimes possible.
One approach that helps is to have a second kernel language,
as explained in Section~\ref{kernellang}.
Another approach is ``loosening and tightening''.
That is, develop the system in semi-independent stages,
where one stage is factored and the next stage brings the factors together.
A typical example is a compiler consisting of
a naive code generator followed by a smart peephole optimizer.

\item It is important to have a sophisticated module system,
with lazy loading, support for mutually-dependent modules,
and support for application deployment.
In Mozart, both Oz and C++ modules can be loaded lazily,
i.e., only when the module is needed.
In this way, the system is both lean and
has lots of functionality.
Lazy loading of Oz modules is implemented with the
\OzInline{ByNeed} operation (see Section~\ref{kernellang}).
Support for mutually-dependent Oz modules means that cyclic
dependencies need to bottom out only at run-time, not at load-time.
This turns out to be important in practice, since modules
often depend on each other.
Support for application deployment 
includes the ability to statically link
a collection of modules into a single module.
This simplifies how modules are offered to users.
A final point is that the module system is written 
within the language, using records, explicit laziness,
and functors implemented by higher-order procedures.

\item It is important to have a powerful interface
to a lower-level language.
Mozart has a C++ interface that allows to add new
constraint systems~\cite{cinterface,cetutorial}.
These constraint systems are fully integrated
into the system, including taking advantage
of the full power of computation spaces.
The current Mozart system has four constraint systems,
based on rational trees
(for both ``bound records'' and ``free records''~\cite{VMS:PLILP96}),
finite domains~\cite{fdtutorial}, and finite sets~\cite{fstutorial}.
Mozart also supports memory management across
the interface, with garbage collection from the Oz side
(using finalization and weak pointers)
and manual control from the C++ side.
\end{itemize}

\subsection{Support true multiparadigm programming}
\label{supportmp}

In any large programming project,
it is almost always a good idea to
use more than one paradigm:
\begin{itemize}
\item Different parts are often best programmed
in different paradigms.\footnote{Another
approach is to use multiple languages with
well-defined interfaces.
This is more complex, but can sometimes work well.}
For example, an event handler may be defined as
an active object whose new state
is a function of its previous state and an external event.
This uses both the object-oriented and functional paradigms
and encapsulates the concurrency in the active object.
\item Different levels of abstraction are often best
expressed in different paradigms.
For example, consider a multi-agent system programmed in
a concurrent logic language.
At the language level, the system does not have the concept of state.
But there is a higher level, the agent level,
consisting of stateful entities called ``agents'' sending
messages to each other.
Strictly speaking,
these concepts do not exist at the language level.
To reason about them, the agent level
is better specified as a graph of active objects.
\end{itemize}
It is always possible to {\em encode} one paradigm
in terms of another.
Usually this is not a good idea.
We explain why in one particularly interesting case,
namely pure concurrent logic programs with state~\cite{portpaper}.
The canonical way to encode state
in a pure concurrent logic program is by using {\em streams}.
An active object is a recursive predicate
that reads an internal stream.
The object's current state is
the internal stream's most-recent element.
A {\em reference} to an active object is a stream that
is read by that object.
This reference can only be used by one sender object,
which sends messages by binding the stream's tail.
Two sender objects sending messages to a third object
are coded as two streams feeding a {\em stream merger},
whose output stream then feeds the third object.
Whenever a new reference is created,
a new stream merger has to be created.
The system as a whole is therefore more complex
than a system with state:
\begin{itemize}
\item The communication graph of the active objects is encoded
as a network of streams and stream mergers.
In this network, each object has a tree of
stream mergers feeding into it.
The trees are created incrementally
during execution, as object references
are passed around the system.
\item To regain efficiency,
the compiler and run-time system must be smart enough to discover
that this network is equivalent to a much simpler structure
in which senders send directly to receivers.
This ``decompilation'' algorithm is so complex
that to our knowledge
no concurrent logic system implements it.
\end{itemize}
On the other hand,
adding state directly to the execution model
makes the system simpler and more uniform.
In that case, programmer-visible state (e.g., active objects with identities) is
mapped directly to execution model state (e.g., using ports for many-to-one
communication),
which is compiled directly into machine state.
Both the compiler and the run-time system are simple.
One may argue that the stateful execution model
is no longer ``pure''.
This is true but irrelevant,
since the stateful model allows simpler reasoning
than the ``pure'' stateless one.

Similar examples can be found for other concepts, e.g.,
higher-orderness,
concurrency,
exception handling,
search,
and
laziness~\cite{book}.
In each case, encoding the concept increases the complexity
of both the program and the system implementation.
In each case, adding the concept to the execution model gives
a simpler and more uniform system.
We conclude that a programming language
should support multiple paradigms.

\subsection{Combine dynamic and static typing}
\label{dynamictype}


We define a {\em type} as a set of values
along with a set of operations on those values.
We say that a language has {\em checked types}
if the system enforces that operations
are only executed with values of correct types.
There are two basic approaches to checked typing,
namely dynamic and static typing.
In {\em static typing}, all variable types are known at compile time.
No type errors can occur at run-time.
In {\em dynamic typing}, the variable type is known with
certainty only when the variable is bound.
If a type error occurs at run-time, then an exception is raised.
Oz is a dynamically-typed language.
Let us examine the trade-offs in each approach.

Dynamic typing puts fewer restrictions
on programs and programming than static typing.
For example, it allows Oz to have an incremental
development environment that is part of the run-time system.
It allows to test programs or program fragments
even when they are in an incomplete or inconsistent state.
It allows truly open programming, i.e., independently-written
components can come together and interact with
as few assumptions as possible about each other.
It allows programs, such as operating systems,
that run indefinitely and grow and evolve.

On the other hand,
static typing has at least three advantages
when compared to dynamic typing.
It allows to catch more program errors at compile time.
It allows for a more efficient implementation,
since the compiler can choose a representation
appropriate for the type.
Last but not least,
it allows for partial program verification,
since some program properties can be guaranteed by the type checker.

In our experience, we find that neither approach is always clearly better.
Sometimes flexibility is what matters; at other times
having guarantees is more important.
It seems therefore that the right type system should be ``mixed'',
that is, be a combination of static and dynamic typing.
This allows the following development methodology,
which is consistent with our experience.
In the early stages of application development,
when we are building prototypes,
dynamic typing is used to maximize flexibility.
Whenever a part of the application is completed,
then it is statically typed
to maximize correctness guarantees and efficiency.
For example, module interfaces and procedure arguments
could be statically typed to maximize early detection of errors.
The most-executed part of a program could be
statically typed to maximize its efficiency.

Much work has been done to
add some of the advantages of dynamic typing to
a statically-typed language, while keeping the
good properties of static typing:
\begin{itemize}
\item Polymorphism adds flexibility to functional and
object-oriented languages.
\item Type inferencing, pioneered by ML, relieves the programmer
of the burden of having to type the whole program explicitly.
\end{itemize}
Our proposal for a mixed type system would go in the opposite direction.
In the mixed type system, the default is dynamic typing.
Static typing is done as soon as needed, but not before.
This means that the trade-off between flexibility and having guarantees
is not frozen by the language design,
but is made available to the programmer.
The design of this mixed type system is a subject for future research.

Mixed typing is related to the concept of ``soft typing'',
an approach to type checking for dynamically-typed languages~\cite{soft}.
In soft typing,
the type checker cannot always decide at compile time
whether the program is correctly typed.
When it cannot decide,
it inserts run-time checks to ensure safe execution.
Mixed typing differs from soft typing in that
we would like to avoid the inefficiency of run-time checking,
which can potentially change a program's time complexity.
The statically-typed parts should be truly statically typed.


\subsection{Use an evolutionary development methodology}
\label{methodology}

The development methodology used in the Oz project
has been refined over many years,
and is largely responsible for the combination
of expressive power, semantic simplicity,
and implementation efficiency in Mozart.
The methodology is nowhere fully described in print;
there are only partial explanations~\cite{opm,sendai99}.
We summarize it here.

At all times during development, there is a robust implementation.
However, the system's design is in continuous flux.
The system's developers continuously introduce
new abstractions as solutions to practical problems.
The burden of proof is on the developer proposing the abstraction:
he must prototype it and show an application
for which it is necessary.
The net effect of a new abstraction
must be either to simplify the system
or to greatly increase its expressive power.
If this seems to be the case,
then intense discussion takes place among all developers 
to simplify the abstraction as much as possible.
Often it vanishes: it can be completely expressed
without modifying the system.
This is not always possible.
Sometimes it is better to modify the system:
to extend it or to replace an existing
abstraction by a new one.

The decision whether to accept an abstraction
is made according to
several criteria including aesthetic ones.
Two major acceptance criteria are related to
implementation and formalization.
The abstraction is acceptable only if
its implementation is efficient
and its formalization is simple.

This methodology extends the approaches put forward by
Hoare, Ritchie, and Thompson~\cite{hoare,ritchie,thompson}.
Hoare advocates designing a program
and its specification concurrently.
He also explains the importance
of having a simple core language.
Ritchie advises having the designers and others actually
use the system during the development period.
In Mozart, as in most Prolog systems, this is
possible because the development environment
is part of the run-time system.
Thompson shows the power of a well-designed abstraction.
The success of Unix was made possible due to its
simple, powerful, and appropriate abstractions.

With respect to traditional software design processes,
this methodology is closest to {\em exploratory programming},
which consists in developing an initial implementation,
exposing it to user comment,
and refining it until the system is adequate~\cite{sommerville}.
The main defect of exploratory programming,
that it results in systems with ill-defined structure,
is avoided by the way the abstractions are refined and
by the double requirement of efficient implementation
and simple formalization.

The two-step process of first generating abstractions and then
selecting among them is analogous to the basic process of evolution.
In evolution, an unending source of different individuals
is followed by a filter, survival of the fittest~\cite{darwin}.
In the analogy, the individuals are abstractions
and the filters are the two acceptance criteria
of efficient implementation and simple formalization.
Some abstractions thrive (e.g., compositionality with lexical scoping),
others die
(e.g., the ``generate and test'' approach to search is dead,
being replaced by propagate and distribute),
others are born and mature
(e.g., dynamic scope, which is currently under discussion),
and others become instances of more general ones
(e.g., deep guards, once basic, are now implemented with spaces).

\section{Conclusions and perspectives}


The Oz language provides powerful tools for both 
the algorithmic and search classes of logic programming problems.
In particular, there are many tools for taming search
in real-world situations. 
These tools include global constraints, search heuristics,
and interactive libraries to visualize and guide the
search process.

Oz is based on a lean execution model that subsumes
deterministic logic programming,
concurrent logic programming,
nondeterministic logic programming,
constraint programming,
strict and nonstrict functional programming,
and concurrent object-oriented programming.
Oz supports declarative concurrency,
a little-known form of concurrent programming
that deserves to be more widely known.
Because of appropriate syntactic and implementation
support, all these paradigms are easy to use. 
We say that Oz is {\em multiparadigm}.
It is important to be multiparadigm
because good program design often requires
different paradigms to be used for different parts of a program.
To a competent Oz programmer, the conventional boundaries
between paradigms are artificial and irrelevant.

The Mozart system implements Oz and is in continuing
development by the Mozart Consortium~\cite{mozart110}.
Research and development started in 1991.
The current release has a full-featured development environment
and is being used for serious application development.
This article covers most of the basic language primitives of Oz.
We only briefly discussed the object system,
the module system (i.e., functors),
and constraint programming, because of space limitations.
In addition to ongoing research in constraint programming,
we are doing research in 
distribution,
fault tolerance,
security,
transactions,
persistence,
programming environments,
software component architectures,
tools for collaborative applications, and
graphic user interfaces.
Another important topic, as yet unexplored,
is the design of a mixed type system
that combines the advantages of static and dynamic typing.
The work on distribution and related areas started in 1995~\cite{perdio}.
Most of these areas are traditionally given short shrift by the
logic and functional programming communities,
yet they merit special attention due to their importance
for real-world applications.

\section*{Acknowledgements}

This article is based on the work of many people over many years.
We thank all the contributors
and developers of the Mozart system.
Many of the opinions expressed are shared
by other members of the Mozart Consortium.
We thank Danny De Schreye for suggesting
the ICLP99 tutorial on which this article is based.
We thank Krzysztof Apt,
Manuel Hermenegildo, Kazunori Ueda,
and others for their questions and comments
at the ICLP99 tutorial
where the original talk was given.
We thank the anonymous referees for their comments
that helped us much improve the presentation.
Finally, we give a special thanks to Juris Reinfelds.
This research was partly financed
by the Walloon Region of Belgium in the PIRATES project.

\bibliography{paper}


\end{document}